\documentclass[12pt]{article}

\usepackage{amsmath}
\usepackage{amsfonts}
\usepackage{color}
\usepackage{latexsym}
\usepackage{fullpage}
\usepackage{tablefootnote}
\usepackage{epsfig}
\usepackage{multirow}
\usepackage{float}
\usepackage{array}
\usepackage{amssymb,amsthm}
\usepackage{hyperref}   
\usepackage{algorithm2e} 
\usepackage{cleveref}    
\usepackage{setspace}
\usepackage{tgtermes} 

\usepackage{mathtools}
\usepackage{mathrsfs}

\usepackage[round]{natbib}
\newcommand*{\R}{\mathbb{R}}

\definecolor{antiquefuchsia}{rgb}{0.57, 0.36, 0.51}
\definecolor{MyViolet}{rgb}{0.45,0.08,0.95}
\definecolor{MyBrown}{rgb}{0.45,0.08,0}
\definecolor{MyDarkBlue}{rgb}{0,0.08,0.45}

\usepackage{makecell,booktabs}
\usepackage[version=4]{mhchem}   
\usepackage[strict]{changepage}

\allowdisplaybreaks

\date{}

\title{Estimation of Spatiotemporal Poisson Processes with Some Missing Location Data}
 
\begin{document}

\include{def}

\maketitle

\begin{center}
\begin{tabular}{ccc}
\begin{tabular}{c}
Vincent Guigues\\
School of Applied Mathematics, FGV\\
Praia de Botafogo, Rio de Janeiro, Brazil\\
{\tt vincent.guigues@fgv.br}
\end{tabular}&
&
\begin{tabular}{c}
Anton J. Kleywegt\\
Georgia Institute of Technology\\
Atlanta, Georgia 30332-0205, USA\\
{\tt anton@isye.gatech.ed}\\
\end{tabular}\\
&&\\
\begin{tabular}{c}
Victor Hugo Nascimento\\
School of Applied Mathematics, FGV\\
Praia de Botafogo, Rio de Janeiro, Brazil\\
{\tt victorhugo.vhrn@gmail.com}
\end{tabular}
&
&
\begin{tabular}{c}
Lucas Rafael de Andrade\\
School of Applied Mathematics, FGV\\
Praia de Botafogo, Rio de Janeiro, Brazil\\
{\tt lucasghostofsparta@hotmail.com}
\end{tabular}
\end{tabular}
\end{center}

\begin{abstract}
We consider models for spatiotemporal Poisson processes with some missing location data.
We discuss four models that make provision for missing location data, and their estimation.
The corresponding code is available on GitHub as an extension of LASPATED at {\url{https://github.com/vguigues/LASPATED/Missing_Data}}.
We tested our models using the process of emergency call arrivals to an emergency medical service where the emergency reports often omit the location of the emergency.
We show the difference made by using models that make provision for missing location data.
\end{abstract}


\section{Introduction}

The study of spatiotemporal data
and the development of statistical models
for such data is important for many real-life applications. One such application is the 
management of a fleet of ambulances
of an Emergency Medical Service (EMS)
which requires studying the process of
emergency calls. These calls
have temporal information: the instant
of the call, spatial information: the location
of the call, and additional features such
as the priority of the call. This application
has been the object of a huge research work,
see for instance
\citep{fitz:73,gree:04,chai:72,jagt:17a,lisay:16,mcla:13b,mcla:13a,band:14,band:12,schm:12,lees:12,lees:11,ande:07,mayo:13,swov:73b,swov:73a,tore:71,dask:83,dask:81,berl:74,chur:74,Talarico2015} and references therein.
For that application, we developed
in \citep{guiklevhn2022}, \citep{ourheuristics23a}
\citep{websiteambrouting24}, optimization,
management, and visualization tools for
the management of the fleet of ambulances
of an EMS. These optimization models
are based on a rolling horizon approach 
\citep{guiguessagastiz2012,guiguessagastiz2010}
where each time a decision is taken a two-stage
stochastic program is solved using
scenarios of future calls. For 
the generation of these scenarios, we developed
statistical models and a library called LASPATED (Library for the Analysis of SPAtio TEmporal
Discrete data), 
see \citep{laspatedpaper, laspatedmanual}
for details. These statistical models assume that
for every call, we both have a time stamp
and the location of the call. 
Our models were tested with data from Rio
de Janeiro EMS.
However, we became aware, collaborating
with this EMS, that many emergency calls came without
the information of the location of the call. In this context,
the objective of this paper is twofold: (i)
develop statistical models for spatiotemporal
data that can deal with missing locations
and test these models with our historical
data of emergency calls to Rio de Janeiro EMS
and (ii) extend LASPATED software incorporating
in LASPATED the
calibration and test functions for the statistical
models we developed to deal with missing
locations.
Though there is already rich literature
on statistical models to deal with missing data, see for instance \citep{bookmd1, bookmd2,bookmd3},
our main contribution is to develop and test
extensions
of the specific models we proposed recently in
LASPATED
\citep{laspatedpaper, laspatedmanual}
to the case where some of the locations are not
present in the data. 
We also make our code
publicly available on GitHub and integrate it
in LASPATED, as an extension of our first version
of LASPATED.
The outline of the paper is the following.
In LASPATED, we proposed two types of models
(one without covariates and one with covariates)
based on a time and space discretization of the
data. If $\mathcal{T}$ is a partition of time
and $\mathcal{I}$ a partition of the studied
area into zones and if $\mathcal{C}$
is the set of arrival types, we assume that
the process of arrivals for time interval
$t \in \mathcal{T}$, zone $i \in \mathcal{I}$,
and arrival type $c \in \mathcal{C}$ is Poisson
with some intensity $\lambda_{c,i,t}$.
In Section \ref{sec:model1}, we propose
a modification of these models
(both non-regularized and regularized) as a splitting of Poisson processes with probabilities of missing the location either fixed or depending on arrival type and time.
In Section \ref{sec:model1m2}, we propose
another extension of the models from LASPATED
where the probability that an observation with missing location comes from a zone is proportional to the population of this zone.
In Section \ref{sec:num}, we present numerical
experiments applied to the aforementioned
data of emergency calls to Rio de Janeiro EMS,
comparing the performance
of our two models with simple estimators that
discard calls without locations. 
\if{
Due to space limitations, figures
of our numerical experiments
are reported in the online companion
\cite{laspatedmissingarxiv} of this
paper and  figure
numbers correspond to the number of 
the figures in this online 
companion.
}\fi

\section{Poisson Model with Estimated Probabilities of Missing Location Data}
\label{sec:model1}

\subsection{Maximum Likelihood Estimators}
\label{sec:model1ml}

Let $\mathcal{I}$ denote the set of zones that form a partition of the region that contains the data, let $\mathcal{T}$ denote a partition of time into time intervals, and let $\mathcal{C}$ denote the set of arrival types.
Let $\mathcal{D}_{t}$ denote the duration of time interval $t \in \mathcal{T}$.
Consider a Poisson process model in which the number of arrivals $X_{c,i,t,n}$ for arrival type $c$, zone $i$, time interval $t$, and observation $n$ is Poisson distributed with mean 
$\lambda_{c,i,t} \mathcal{D}_{t}$.
For every arrival type $c$ and time interval $t$, let
\begin{equation}
\label{sct}
S_{c,t} \ \ := \ \ \sum_{i \in \mathcal{I}} \lambda_{c,i,t}
\end{equation}
denote the sum of Poisson intensities over all zones.
Assume that for each arrival, there is a probability~$p$ that the location of the arrival is not reported (and probability $1-p$ that the location is reported).
The Bernoulli random variables indicating whether the location is reported or not are independent.
In this model, the probability of reporting the location does not depend on the zone, the arrival type, or the time interval.

Let $\mathcal{A}$ denote the set of arrivals, which is partitioned into $2$~subsets: $\mathcal{A} = \mathcal{A}^{0} \cup \mathcal{A}^{1}$.
The arrivals in $\mathcal{A}^{0}$ do not contain data about the location of the arrival, and the arrivals in $\mathcal{A}^{1}$ contain data about the location of the arrival.

Observation refers to a count of the number of arrivals in $\mathcal{A}^{0}$ for an arrival type~$c$ during a time interval~$t$ or to a count of the number of arrivals in $\mathcal{A}^{1}$ for an arrival type~$c$ in a zone~$i$ during a time interval~$t$.
For each $c \in \mathcal{C}$ and $t \in \mathcal{T}$, let $N^{0}_{c,t}$ denote the number of observations in $\mathcal{A}^{0}$ for arrival type~$c$ and time interval~$t$, and let these observations be indexed $n = 1,\ldots,N^{0}_{c,t}$.
For each $c \in \mathcal{C}$, $i \in \mathcal{I}$, and $t \in \mathcal{T}$, let $N^{1}_{c,i,t}$ denote the number of observations for arrival type~$c$, zone~$i$, and time interval~$t$, and let these observations be indexed $n = 1,\ldots,N^{1}_{c,i,t}$.
Assume that for each $c \in \mathcal{C}$, $i \in \mathcal{I}$, and $t \in \mathcal{T}$, it holds that $N^{1}_{c,i,t} = N^{0}_{c,t}$, and that observation $n \in \{1,\ldots,N^{1}_{c,i,t}\}$ corresponds to observation $n \in \{1,\ldots,N^{0}_{c,t}\}$, for example, these observations were made during the same week.
In what follows, we use the simplified notation $N_{c,t}$ for the common value of $N_{c,t}^{0}$ and $N_{c,i,t}^{1}$.

The Poisson process of all arrivals is the aggregation of the following two independent Poisson processes: 
\begin{itemize}
\item
the Poisson process of arrivals for which the location is reported, with corresponding random number of arrivals $Y_{c,i,t,n}$ for arrival type~$c$, zone~$i$, time interval~$t$, and observation $n$, and
\item
the Poisson process of arrivals for which the location is not reported with corresponding random number of arrivals $Z_{c,t,n}$ for arrival type~$c$, time interval~$t$, and observation~$n$.
\end{itemize}
Thus $Y_{c,i,t,n}$ is Poisson with mean $(1-p) \lambda_{c,i,t} \mathcal{D}_{t}$, and $Z_{c,t,n}$ is Poisson with mean $p S_{c,t} \mathcal{D}_{t}$.
For the process of calls with reported locations, let $M_{c,i,t,n}^{1}$ denote the number of arrivals for arrival type~$c$, zone~$i$, time interval~$t$, and observation~$n$.
Also, let 
\[
M_{c,\bullet,t,n}^{1} \ \ := \ \ \sum_{i \in \mathcal{I}} M_{c,i,t,n}^{1}
\]
denote the total number of arrivals over all zones for which the zone is reported, for arrival type~$c$, time interval~$t$, and observation~$n$, let
\[
M_{c,i,t,\bullet}^{1} \ \ := \ \ \sum_{n=1}^{N_{c,t}}  M_{c,i,t,n}^{1}
\]
denote the total number of arrivals over all observations for which the zone is reported, for arrival type~$c$, zone~$i$, and time interval~$t$, let
\[
M_{c,\bullet,t,\bullet}^{1} \ \ := \ \ \sum_{i \in \mathcal{I}} \sum_{n=1}^{N_{c,t}} M_{c,i,t,n}^{1}
\]
denote the total number of arrivals over all zones and observations for which the zone is reported, for arrival type~$c$, time interval~$t$, and let
\[
M^{1} \ \ := \ \ \sum_{c \in \mathcal{C}} \sum_{i \in \mathcal{I}} \sum_{t \in \mathcal{T}} \sum_{n=1}^{N_{c,t}} M_{c,i,t,n}^{1}
\]
denote the total number of all arrivals for which the location is reported.

For the process of calls with locations not reported, let $M_{c,t,n}^{0}$ denote the number of arrivals for arrival type~$c$, time interval~$t$, and observation~$n$.
Also, let
\[
M_{c,t,\bullet}^{0} \ \ := \ \ \sum_{n=1}^{N_{c,t}} M_{c,t,n}^{0}
\]
denote the total number of arrivals over all observations for which the zone is not reported, for arrival type~$c$ and time interval~$t$, and let
\[
M^{0} \ \ := \ \ \sum_{c \in \mathcal{C}} \sum_{t \in \mathcal{T}} \sum_{n=1}^{N_{c,t}} M_{c,t,n}^{0}
\]
the total number of all arrivals for which the zone is not reported.

Next we consider estimation of parameters $\lambda := (\lambda_{c,i,t}, c \in \mathcal{C}, i \in \mathcal{I}, t \in \mathcal{T})$ and $p$.

The likelihood function of the observed data is given by
\begin{align*}
L_{1}(\lambda,p) \ \ & = \ \ \prod_{c \in \mathcal{C}} \prod_{t \in \mathcal{T}} \prod_{n=1}^{N_{c,t}} \mathbb{P}(Z_{c,t,n} = M_{c,t,n}^{0}) \prod_{i \in \mathcal{I}} \mathbb{P}(Y_{c,i,t,n} = M_{c,i,t,n}^{1}) \\
& = \ \ \prod_{c \in \mathcal{C}} \prod_{t \in \mathcal{T}} \prod_{n=1}^{N_{c,t}} \exp(-p S_{c,t} \mathcal{D}_{t}) \frac{(p S_{c,t}\mathcal{D}_{t})^{M_{c,t,n}^{0}}}{M_{c,t,n}^{0}!} \prod_{i \in \mathcal{I}} \exp(-(1-p) \lambda_{c,i,t} \mathcal{D}_{t}) \frac{((1-p) \lambda_{c,i,t} \mathcal{D}_{t})^{M_{c,i,t,n}^{1}}}{M_{c,i,t,n}^{1}!}
\end{align*}
and therefore the log-likelihood function is given, up to a constant term that can be dropped for maximization, by
\[ 
\mathscr{L}_{1}(\lambda,p) \ \ = \ \ M^{1} \log(1-p) + M^{0} \log(p) + \sum_{c \in \mathcal{C}} \sum_{t \in \mathcal{T}} \left(-N_{c,t} S_{c,t} \mathcal{D}_{t} + M_{c,t,\bullet}^{0} \log(S_{c,t}) + \sum_{i \in \mathcal{I}} M_{c,i,t,\bullet}^{1} \log(\lambda_{c,i,t})\right).  
\]
The estimator $(\hat{\lambda},\hat{p})$ that maximizes $\mathscr{L}_{1}(\lambda,p)$ satisfies
\[
\frac{\partial \mathscr{L}_{1}(\hat{\lambda},\hat{p})}{\partial \lambda} \ \ = \ \ 0 \quad \mbox{and} \quad \frac{\partial \mathscr{L}_{1}(\hat{\lambda},\hat{p})}{\partial p} \ \ = \ \ 0
\]
which gives
\begin{equation}
\label{solvelp1}
\begin{array}{l}
\displaystyle \frac{M^{0}}{\hat{p}} - \frac{M^{1}}{1 - \hat{p}} \ \ = \ \ 0 \\
\displaystyle \frac{M_{c,t,\bullet}^{0}}{\hat{S}_{c,t}} - N_{c,t} \mathcal{D}_{t} + \frac{M_{c,i,t,\bullet}^{1}}{\hat{\lambda}_{c,i,t}} \ \ = \ \ 0, \quad \forall \ c \in \mathcal{C}, \ t \in \mathcal{T}, \ i \in \mathcal{I},
\end{array}
\end{equation}
where
\begin{equation}
\label{defshat}
\hat{S}_{c,t} \ \ = \ \ \sum_{i \in \mathcal{I}} \hat{\lambda}_{c,i,t}.
\end{equation}
It follows from~\eqref{solvelp1} that
\begin{equation}
\label{formp}
\hat{p} \ \ = \ \ \frac{M^{0}}{M^{1} + M^{0}}.
\end{equation}
Thus \eqref{formp} provides a very natural estimator of $p$ which is the empirical probability of not reporting the location, namely the total number $M^{0}$ of arrivals where the location is not reported divided by the total number $M^{1} + M^{0}$ of arrivals.
It also follows from~\eqref{solvelp1} that
\begin{equation}
\label{lambdaf1}
\hat{S}_{c,t} \ \ = \ \ \frac{M_{c,\bullet,t,\bullet}^{1} + M_{c,t,\bullet}^{0}}{N_{c,t} \mathcal{D}_{t}}.
\end{equation}
and
\begin{equation}
\label{lambdaf}
\hat{\lambda}_{c,i,t} \ \ = \ \ \frac{\hat{S}_{c,t} M_{c,i,t,\bullet}^{1}}{M_{c,\bullet,t,\bullet}^{1}}
\ \ = \ \ \underbrace{\left(\frac{M_{c,\bullet,t,\bullet}^{1} + M_{c,t,\bullet}^{0}}{M_{c,\bullet,t,\bullet}^{1}}\right)}_{1 / (1 - \hat{p}_{c,t})}
\underbrace{\left(\frac{M_{c,i,t,\bullet}^{1}}{N_{c,t} \mathcal{D}_{t}}\right)}_{\overline{\lambda}_{c,i,t} / \mathcal{D}_{t}}.
\end{equation}
Thus a very natural empirical estimator of intensities $\lambda_{c,i,t}$ is obtained.
Here,
\begin{equation}
\label{estimpct}
\hat{p}_{c,t} \ \ = \ \ \frac{M_{c,t,\bullet}^{0}}{M_{c,\bullet,t,\bullet}^{1} + M_{c,t,\bullet}^{0}}.
\end{equation}
denotes the empirical estimator of the probability~$p$ of not reporting for arrival type~$c$ and time interval~$t$, and
\[
\overline{\lambda}_{c,i,t} \ \ = \ \ \frac{M_{c,i,t,\bullet}^{1}}{N_{c,t}}
\]
denotes the empirical estimator of the mean number of arrivals $(1-p) \lambda_{c,i,t} \mathcal{D}_{t}$, for arrival type~$c$, zone~$i$, and time interval~$t$, for which the location is reported.

Next consider a model in which the probabilities $p_{c,t}$ of not reporting the location depend on arrival type~$c$ and time period~$t$.
Let $p := (p_{c,t}, c \in \mathcal{C}, t \in \mathcal{T})$.
Now $Y_{c,i,t,n}$ is Poisson with mean $(1-p_{c,t}) \lambda_{c,i,t} \mathcal{D}_{t}$ and $Z_{c,t,n}$ is Poisson with mean $p_{c,t} S_{c,t} \mathcal{D}_{t}$.

The likelihood of the model is given by
\begin{align*}
& L_{2}(p,\lambda)    = \prod_{c \in \mathcal{C}} \prod_{t \in \mathcal{T}} \prod_{n=1}^{N_{c,t}} \mathbb{P}(Z_{c,t,n} = M_{c,t,n}^{0}) \prod_{i \in \mathcal{I}} \mathbb{P}(Y_{c,i,t,n} = M_{c,i,t,n}^{1}) \\
& =  \prod_{c \in \mathcal{C}} \prod_{t \in \mathcal{T}} \prod_{n=1}^{N_{c,t}} \exp(-p_{c,t} S_{c,t} \mathcal{D}_{t}) \frac{(p_{c,t} S_{c,t} \mathcal{D}_{t})^{M_{c,t,n}^{0}}}{M_{c,t,n}^{0}!} \prod_{i \in \mathcal{I}} \exp(-(1-p_{c,t}) \lambda_{c,i,t} \mathcal{D}_{t}) \frac{((1-p_{c,t}) \lambda_{c,i,t} \mathcal{D}_{t})^{M_{c,i,t,n}^{1}}}{M_{c,i,t,n}^{1}!}
\end{align*}
and therefore the log likelihood is given, up to a constant term that can be dropped for maximization, by
\begin{align*}
\mathscr{L}_{2}(\lambda,p) \ \ = \ \ & \sum_{c \in \mathcal{C}} \sum_{t \in \mathcal{T}} \left(-N_{c,t} S_{c,t} \mathcal{D}_{t} + M_{c,t,\bullet}^{0} \log(S_{c,t}) + \sum_{i \in \mathcal{I}} M_{c,i,t,\bullet}^{1} \log(\lambda_{c,i,t})\right) \\
& {} + \sum_{c \in \mathcal{C}} \sum_{t \in \mathcal{T}} \Big(M_{c,t,\bullet}^{0} \log(p_{c,t}) + M_{c,\bullet,t,\bullet}^{1} \log(1-p_{c,t})\Big).
\end{align*}
The pair $(\hat{\lambda},\hat{p})$ maximizing $\mathscr{L}_{2}(\lambda,p)$ solves
\[
\frac{\partial \mathscr{L}_{2}(\hat{\lambda},\hat{p})}{\partial \lambda} \ \ = \ \ 0 \quad \mbox{and} \quad \frac{\partial \mathscr{L}_{2}(\hat{\lambda},\hat{p})}{\partial p} \ \ = \ \ 0
\]
which gives
\begin{equation}
\label{solvelp}
\begin{array}{l}
\displaystyle \frac{M_{c,t,\bullet}^{0}}{\hat{p}_{c,t}} - \frac{M_{c,\bullet,t,\bullet}^{1}}{1-\hat{p}_{c,t}} \ \ = \ \ 0, \quad \forall \ c \in \mathcal{C}, \ t \in \mathcal{T} \\
\displaystyle \frac{M_{c,t,\bullet}^{0}}{\hat{S}_{c,t}} - N_{c,t} \mathcal{D}_{t} + \frac{M_{c,i,t,\bullet}^{1}}{\hat{\lambda}_{c,i,t}} \ \ = \ \ 0, \quad \forall \ c \in \mathcal{C}, \ i \in \mathcal{I}, \ t \in \mathcal{T},
\end{array}
\end{equation}
where $\hat{S}_{c,t}$ is given by~\eqref{defshat}.
Thus,
\begin{equation}
\label{estimpct}
\hat{p}_{c,t} \ \ = \ \ \frac{M_{c,t,\bullet}^{0}}{M_{c,\bullet,t,\bullet}^{1} + M_{c,t,\bullet}^{0}}.
\end{equation}
and $\hat{\lambda}_{c,i,t}$ is still given by~\eqref{lambdaf}, as for the previous model.

Observe that when $\hat{p}_{c,t} = 0$ (that is, zones are reported for all arrivals), then $\hat{\lambda}_{c,i,t}$ equals the usual maximum likelihood estimator of a Poisson intensity, although $\mathscr{L}_{2}$ is not defined if $\hat{p}_{c,t} = 0$.
Also, $\hat{\lambda}_{c,i,t}$ cannot be estimated if $M_{c,\bullet,t,\bullet}^{1} = 0$, that is, if $\hat{p}_{c,t} = 1$.

To derive asymptotic confidence intervals for the estimators, we compute the Fisher information matrix of the models.
For the model with log-likelihood $\mathscr{L}_{2}$ and parameter $\theta := (\lambda,p)$, the Fisher information matrix has entries $-\mathbb{E}\left[\frac{\partial^2 \mathscr{L}_{2}(\theta)}{\partial \theta_{i} \partial \theta_{j}}\right]$.
Note that
\begin{align*}
-\frac{\partial \mathscr{L}_{2}(\lambda,p)}{\partial p_{c,t}} \ \ & = \ \ -\frac{M_{c,t,\bullet}^{0}}{p_{c,t}}+\frac{M_{c,\bullet,t,\bullet}^{1}}{1-p_{c,t}} \qquad \forall \ c \in \mathcal{C}, \ t \in \mathcal{T} \\
-\frac{\partial \mathscr{L}_{2}(\lambda,p)}{\partial \lambda_{c,i,t}} \ \ & = \ \ -\frac{M_{c,t,\bullet}^{0}}{S_{c,t}} + N_{c,t} \mathcal{D}_{t} - \frac{M_{c,i,t,\bullet}^{1}}{\lambda_{c,i,t}} \qquad \forall \ c \in \mathcal{C}, \ i \in \mathcal{I}, \ t \in \mathcal{T}.
\end{align*}
The second-order derivatives are therefore given by
\begin{align*}
-\frac{\partial^2 \mathscr{L}_{2}(\lambda,p)}{\partial p_{c,t}^2} \ \ & = \ \ \frac{M_{c,t,\bullet}^{0}}{p_{c,t}^2} + \frac{M_{c,\bullet,t,\bullet}^{1}}{(1-p_{c,t})^2} \qquad \forall \ c \in \mathcal{C}, \ t \in \mathcal{T} \\
-\frac{\partial^2 \mathscr{L}_{2}(\lambda,p)}{\partial \lambda_{c,i,t}^2} \ \ & = \ \ \frac{M_{c,t,\bullet}^{0}}{S_{c,t}^2} + \frac{M_{c,i,t,\bullet}^{1}}{\lambda_{c,i,t}^2} \qquad \forall \ c \in \mathcal{C}, \ i \in \mathcal{I}, \ t \in \mathcal{T} \\
-\frac{\partial^2 \mathscr{L}_{2}(\lambda,p)}{\partial \lambda_{c,i,t} \partial \lambda_{c,j,t}} \ \ & = \ \ \frac{M_{c,t,\bullet}^{0}}{S_{c,t}^2} \qquad \forall \ c \in \mathcal{C}, \ i,j \in \mathcal{I} \mbox{ with } i \neq j, \ t \in \mathcal{T} \\
-\frac{\partial^2 \mathscr{L}_{2}(\lambda,p)}{\partial \lambda_{c,i,t} \partial \lambda_{c',j,t'}} \ \ & = \ \ 0 \qquad \mbox{if } c \neq c' \mbox{ or } t \neq t' \\
-\frac{\partial^2 \mathscr{L}_{2}(\lambda,p)}{\partial p_{c',t'} \partial \lambda_{c,i,t}} \ \ & = \ \ 0 \qquad \forall \ c,c' \in \mathcal{C}, \ i \in \mathcal{I}, \ t,t' \in \mathcal{T} \\
-\frac{\partial^2 \mathscr{L}_{2}(\lambda,p)}{\partial p_{c,t} \partial p_{c',t'}} \ \ & = \ \ 0 \qquad \mbox{if } c \neq c' \mbox{ or } t \neq t'.
\end{align*}
Also,
\begin{align*}
\mathbb{E}[M_{c,t,\bullet}^{0}] \ \ & = \ \ p_{c,t}\mathcal{D}_{t} N_{c,t}S_{c,t} \\
\mathbb{E}[M_{c,\bullet,t,\bullet}^{1}] \ \ & = \ \ (1-p_{c,t})\mathcal{D}_{t} |\mathcal{I}|N_{c,t}\lambda_{c,i,t} \\
\mathbb{E}[M_{c,i,t,\bullet}^{1}] \ \ & = \ \ (1-p_{c,t})\mathcal{D}_{t} N_{c,t}\lambda_{c,i,t}.
\end{align*}
Thus the entries of the Fisher information matrix are given by
\begin{align*}
& \mathbb{E}\left[-\frac{\partial^2 \mathscr{L}_{2}(\lambda,p)}{\partial p_{c,t}^2}\right] \ \ = \ \ \frac{\mathcal{D}_{t} N_{c,t} S_{c,t}}{p_{c,t}} + \frac{\mathcal{D}_{t} |\mathcal{I}| N_{c,t} \lambda_{c,i,t}}{1-p_{c,t}} \qquad \forall \ c \in \mathcal{C}, \ t \in \mathcal{T} \\
& \mathbb{E}\left[-\frac{\partial^2 \mathscr{L}_{2}(\lambda,p)}{\partial \lambda_{c,i,t}^2}\right] \ \ = \ \ \frac{p_{c,t} \mathcal{D}_{t} N_{c,t}}{S_{c,t}} + \frac{(1-p_{c,t}) \mathcal{D}_{t} N_{c,t}}{ \lambda_{c,i,t}} \qquad \forall \ c \in \mathcal{C}, \ i \in \mathcal{I}, \ t \in \mathcal{T}\\
& \mathbb{E}\left[-\frac{\partial^2 \mathscr{L}_{2}(\lambda,p)}{\partial \lambda_{c,i,t} \partial \lambda_{c,j,t}}\right] \ \ = \ \ \frac{p_{c,t}\mathcal{D}_{t} N_{c,t}}{S_{c,t}} \qquad \forall \ c \in \mathcal{C}, \ i, j \in \mathcal{I} \mbox{ with } i \neq j, \ t \in \mathcal{T}.
\end{align*}
Variance estimates of the maximum likelihood estimators $\hat{\lambda}_{c,i,t}$ and $\hat{p}_{c,t}$ of $\lambda_{c,i,t}$ and $p_{c,t}$ are found on the diagonal of the inverse $I(\theta)^{-1}$ of the Fisher information matrix $I(\theta)$.
For example, with $\hat{V}^{\lambda}_{c,i,t}$ denoting the estimated variance of $\hat{\lambda}_{c,i,t}$, an asymptotic $1-\alpha$ confidence interval for $\lambda_{c,i,t}$ is given by
$$
\left[\hat{\lambda}_{c,i,t} - \Phi^{-1}(1-\alpha/2) \sqrt{\hat{V}^{\lambda}_{c,i,t}}, \ \hat{\lambda}_{c,i,t} + \Phi^{-1}(1-\alpha/2) \sqrt{\hat{V}^{\lambda}_{c,i,t}}\right]
$$
where $\Phi^{-1}(1-\alpha/2)$ denotes the $1-\alpha/2$ quantile of the standard normal distribution.
Similarly, a confidence interval for $p_{c,t}$ can be derived from $\hat{p}_{c,t}$ and $I(\theta)^{-1}$.

\subsection{Regularized Estimators}
\label{sec:reg}

In this section we consider regularized estimators.
Based on information about the application, some groups of time periods may be expected to have intensities that are close to each other.
For example, the intensities for Tuesday [10:00,11:00] and Thursday [10:00,11:00] may be expected to be similar.
To use such information, $\mathcal{T}$ is partitioned into a collection $\mathcal{G}$ of subsets of $\mathcal{T}$ in such a way that it is expected that, for each $c \in \mathcal{C}$, $i \in \mathcal{I}$, and $G \in \mathcal{G}$, the values of $\lambda_{c,i,t}$ for different $t \in G$ will be close to each other (but not necessarily the same), and for each $c \in \mathcal{C}$ and $G \in \mathcal{G}$, the values $p_{c,t}$ for different $t \in G$ will be close to each other.
(Different partitions $\mathcal{G}$ may be used for different $c$ and $i$, but the notation here does not show such dependence.)
Let $W_{G} \ge 0$ denote a similarity weight for $G \in \mathcal{G}$.

Similarly, some groups of zones may be expected to have intensities that are close to each other.
Such an idea can also be incorporated by partitioning the set $\mathcal{I}$ of zones into subsets, each with its own similarity weight, or by specifying for each pair $i,j \in \mathcal{I}$ a similarity weight $w_{i,j} \ge 0$.
An example loss function with similarity regularization by time and space is
\begin{align}
\ell(\lambda,p) \ \ = \ \ & -\mathscr{L}_{2}(\lambda,p) + \sum_{c \in \mathcal{C}} \sum_{i \in \mathcal{I}} \sum_{G \in \mathcal{G}} \frac{W_{G}}{2} \sum_{t,t' \in G} N_{c,t} N_{c,t'} \left(\lambda_{c,i,t} - \lambda_{c,i,t'}\right)^2 \nonumber \\
\label{eqn:regularization1}
& {} + \sum_{c \in \mathcal{C}} \sum_{i,j \in \mathcal{I}} \sum_{t \in \mathcal{T}} \frac{w_{i,j}}{2} N_{c,t}^2 \left(\lambda_{c,i,t} - \lambda_{c,j,t}\right)^2 \\
& {} + \sum_{c \in \mathcal{C}} \sum_{G \in \mathcal{G}} \frac{W_{G}}{2} \sum_{t,t' \in G} N_{c,t} N_{c,t'} \left(p_{c,t} - p_{c,t'}\right)^2. \nonumber
\end{align}
Estimates of $\lambda_{c,i,t}$ and $p_{c,t}$ are then given by solutions of the optimization problem
\begin{equation}
\label{pboptinit}
\min_{\lambda \geq 0, \; 0 \leq p_{c,t} \leq 1} \; \ell(\lambda,p).
\end{equation}
As before, problem \eqref{pboptinit} is separable and can be written
\begin{equation}
\label{pboptinit1}
\min_{\lambda \geq 0} \; \ell_{1}(\lambda) \ + \ \min_{0 \leq p_{c,t} \leq 1} \; \ell_{2}(p)
\end{equation}
where
\begin{align*}
\ell_{1}(\lambda) \ \ & = \ \ \sum_{c \in \mathcal{C}} \sum_{t \in \mathcal{T}} \left(N_{c,t} S_{c,t} \mathcal{D}_{t} - M_{c,t,\bullet}^{0} \log(S_{c,t}) - \sum_{i \in \mathcal{I}}
M_{c,i,t,\bullet}^{1} \log(\lambda_{c,i,t})\right) \\
& \hspace{10mm} {} + \sum_{c \in \mathcal{C}} \sum_{i \in \mathcal{I}} \sum_{G \in \mathcal{G}} \frac{W_{G}}{2} \sum_{t,t' \in G} N_{c,t} N_{c,t'} \left(\lambda_{c,i,t} - \lambda_{c,i,t'}\right)^2 + \sum_{c \in \mathcal{C}} \sum_{i,j \in \mathcal{I}} \sum_{t \in \mathcal{T}} \frac{w_{i,j}}{2} N_{c,t}^2 \left(\lambda_{c,i,t} - \lambda_{c,j,t}\right)^2, \\
\ell_{2}(p) \ \ & = \ \ - \sum_{c \in \mathcal{C}} \sum_{t \in \mathcal{T}} \Big(M_{c,t,\bullet}^{0} \log(p_{c,t}) + M_{c,\bullet,t,\bullet}^{1}\log(1-p_{c,t})\Big) + \sum_{c \in \mathcal{C}} \sum_{G \in \mathcal{G}} \frac{W_{G}}{2} \sum_{t,t' \in G} N_{c,t} N_{c,t'} \left(p_{c,t} - p_{c,t'}\right)^2.
\end{align*}

\subsection{Poisson Model with Arrival Rates Given by Covariates}
\label{sec:modelcov}

Sometimes additional data are available that can provide improved estimates.
For example, each type~$c$, zone~$i$, and time interval~$t$ may have covariates that are correlated with the arrival rates~$\lambda_{c,i,t}$.
For each $c \in \mathcal{C}$, $i \in \mathcal{I}$, and $t \in \mathcal{T}$, let $x_{c,i,t} = (x_{c,i,t,1},\ldots,x_{c,i,t,K})$ denote the covariate values of type~$c$, zone~$i$, and time interval~$t$.
For example, $x_{c,i,t,1}$ may be the population count with home addresses in zone~$i$, and $x_{c,i,t,2}$ may be an indicator that a major event is scheduled in zone~$i$ during time interval~$t$.
Then consider the model
\begin{equation}
\label{eqn:covariate model}
\lambda(x_{c,i,t}) \mathcal{D}_{t} \ \ = \ \ \beta^{\top} x_{c,i,t}
\end{equation}
where $\beta = (\beta_{1},\ldots,\beta_{K})$ are the model parameters.
Let $\mathcal{X}_{c,i,t}$ denote the set of all possible values of $x_{c,i,t}$.
Note that it should hold that
\begin{equation}
\beta^{\top} x_{c,i,t} \ \ \ge \ \ 0 \qquad \forall \ x_{c,i,t} \in \mathcal{X}_{c,i,t}, \ \forall \ c \in \mathcal{C}, i \in \mathcal{I}, t \in \mathcal{T}.
\label{eqn:feassetref1}
\end{equation}
For the random variables $Y_{c,i,t,n}$ and $Z_{c,t,n}$ defined previously, in the model with covariates $Y_{c,i,t,n}$ is Poisson with mean $(1-p_{c,t}) \lambda_{c,i,t} \mathcal{D}_{t} = (1-p_{c,t}) \beta^{\top} x_{c,i,t}$ and $Z_{c,t,n}$ is Poisson with mean $p_{c,t} S_{c,t} \mathcal{D}_{t} = p_{c,t} \beta^{\top} \left(\sum_{i \in \mathcal{I}} x_{c,i,t}\right)$.

Next we give an example of such a model for which numerical results are given in Section~\ref{sec:num}.
Let $\mathcal{T} = \{1,\ldots,48\}$ denote the set of indices of $30$-minute intervals during a day.
Let $\mathcal{D}$ denote the set of indices of the normal days of the week, as well as separate indices for special days such as holidays.
Thus, the cardinality of $\mathcal{D}$ is 7 plus the number of special days.
A pair $(d,t) \in \mathcal{D} \times \mathcal{T}$ specifies a time interval.
Let $\mathcal{O}$ denote the set of occupational land uses.
For each zone $i \in \mathcal{I}$, the covariates are $x_{i} \in \mathbb{R}^{1+|\mathcal{O}|}$, where $x_{i}(1)$ denotes the population count in zone~$i$, and $x_{i}(j+1)$ is the area (in km$^2$) of occupational land use~$j$ in zone~$i$, for $j=1,\ldots,|\mathcal{O}|$.
We assume that these data (population and land type areas) do not depend on time, which is reasonable for moderate time periods.
Then, for each type~$c$, zone~$i$, day~$d$, and daily time period~$t$, the arrival intensity is given by $\lambda_{c,i,d,t} \mathcal{D}_{t} = \beta_{c,d,t}^{\top} x_{i}$ for some $\beta_{c,d,t} \in \mathbb{R}^{1+|\mathcal{O}|}$).
The likelihood for the model is given by
\[
L_{3}(p,\beta) \ \ = \ \ \prod_{c \in \mathcal{C}} \prod_{d \in \mathcal{D}} \prod_{t \in \mathcal{T}} \prod_{n=1}^{N_{c,t}} \mathbb{P}(Z_{c,t,n} = M_{c,d,t,n}^{0}) \prod_{i \in \mathcal{I}} \mathbb{P}(Y_{c,i,t,n} = M_{c,i,d,t,n}^{1})
\]
where
\begin{align*}
& \mathbb{P}(Z_{c,t,n} = M_{c,d,t,n}^{0}) \ \ = \ \ \exp\left(-p_{c,d,t} \beta_{c,d,t}^{\top} \sum_{i \in \mathcal{I}} x_{i}\right) \frac{\left(p_{c,d,t} \beta_{c,d,t}^{\top} \left(\sum_{i \in \mathcal{I}} x_{i}\right)\right)^{M_{c,d,t,n}^{0}}}{M_{c,d,t,n}^{0}!} \\
& \mathbb{P}(Y_{c,i,t,n} = M_{c,i,d,t,n}^{1}) \ \ = \ \ \exp\left(-(1-p_{c,d,t})\beta_{c,d,t}^{\top} x_{i}\right) \frac{\left((1-p_{c,d,t}) \beta_{c,d,t}^{\top} x_{i}\right)^{M_{c,i,d,t,n}^{1}}}{M_{c,i,d,t,n}^{1}!}.
\end{align*}
Therefore the log likelihood is given, up to a constant term that we can drop for maximization, by
\begin{align*}
\mathscr{L}_{3}(\beta,p) \ \ = \ \ & \sum_{c \in \mathcal{C}} \sum_{d \in \mathcal{D}} \sum_{t \in \mathcal{T}} \left(-N_{c,d,t} \beta_{c,d,t}^{\top} \sum_{i \in \mathcal{I}} x_{i} + M_{c,d,t,\bullet}^{0} \log\left(\beta_{c,d,t}^{\top} \sum_{i \in \mathcal{I}} x_{i}\right) + \sum_{i \in \mathcal{I}} M_{c,i,d,t,\bullet}^{1} \log\left(\beta_{c,d,t}^{\top} x_{i}\right)\right) \\
& {} + \sum_{c \in \mathcal{C}} \sum_{d \in \mathcal{D}} \sum_{t \in \mathcal{T}} \Big(M_{c,d,t,\bullet}^{0} \log(p_{c,d,t}) + M_{c,\bullet,d,t,\bullet}^{1} \log(1-p_{c,d,t})\Big),
\end{align*}
where $M_{c,\bullet,d,t,\bullet}^{1} := \sum_{i \in \mathcal{I}} M_{c,i,d,t,\bullet}^{1}$.
The resulting estimator of $p_{c,d,t}$ is given by
\[
\hat{p}_{c,d,t} \ \ = \ \ \frac{M_{c,d,t,\bullet}^{0}}{M_{c,\bullet,d,t,\bullet}^{1}+M_{c,d,t,\bullet}^{0}}
\qquad \forall \ c \in \mathcal{C}, \ d \in \mathcal{D}, \ t \in \mathcal{T}.
\]
Therefore, the estimation problem to find $\hat{\beta}_{c,d,t}$ is
\begin{equation}
\label{model2}
\begin{array}{rl}
\min & \displaystyle \sum_{c \in \mathcal{C}} \sum_{d \in \mathcal{D}} \sum_{t \in \mathcal{T}} \left(N_{c,d,t} \beta_{c,d,t}^{\top} \sum_{i \in \mathcal{I}} x_{i} - M_{c,d,t,\bullet}^{0} \log\left(\beta_{c,d,t}^{\top} \sum_{i \in \mathcal{I}} x_{i}\right) - \sum_{i \in \mathcal{I}} M_{c,i,d,t,\bullet}^{1} \log\left(\beta_{c,d,t}^{\top} x_{i}\right)\right) \\
\mbox{s.t.} & \ \ \beta_{c,d,t}^{\top} x_{i} \ \geq 0,
\quad 0 \ \leq \ \beta_{c,d,t}(1) \ \leq \ 1,
\qquad \forall \ c \in \mathcal{C}, \ d \in \mathcal{D}, \ t \in \mathcal{T}.
\end{array}
\end{equation}

\section{Poisson Model with Location Probabilities Given by Covariates}
\label{sec:model1m2}

In this section we describe a model in which the probability $\pi_{i}$ that an observation with missing location has location~$i$ is given by $\pi_{i} = P_{i} / P$, where $P_{i}$ is the population of zone $i$ and $P$ is the total population of all zones.
As before, the number of arrivals for arrival type~$c$, zone~$i$, and time interval~$t$, is Poisson with mean $\lambda_{c,i,t} \mathcal{D}_{t}$.
For any $\bar{M} \in \mathbb{N}$, let 
\[
\mathcal{N}(\bar{M}) \ \ := \ \ \left\{M = (M_{i}, i \in \mathcal{I}) \; : \; \sum_{i \in \mathcal{I}} M_{i} = \bar{M}, \; M_{i} \in \mathbb{N} \; \forall \; i \in \mathcal{I}\right\}.
\]
Using the same notation for the data as before, the likelihood of this model is given by
\[
L_{4}(\lambda) \ \ = \ \ \prod_{c \in \mathcal{C}} \prod_{t \in \mathcal{T}} \prod_{n=1}^{N_{c,t}} \sum_{M \in \mathcal{N}(M^{0}_{c,t,n})} M^{0}_{c,t,n}! \left[\prod_{i \in \mathcal{I}} \frac{\pi_{i}^{M_{i}}}{M_{i}!}\right] \left[\prod_{i \in \mathcal{I}} \exp\left(-\lambda_{c,i,t} \mathcal{D}_{t}\right) \frac{\left(\lambda_{c,i,t} \mathcal{D}_{t}\right)^{M_{i} + M^{1}_{c,i,t,n}}}{(M_{i} + M^{1}_{c,i,t,n})!}\right].
\]
Therefore the log-likelihood is given by
\begin{align}
\label{eqn:maxlikelihood4}
\mathscr{L}_{4}(\lambda) \ \ & = \ \ \sum_{c \in \mathcal{C}} \sum_{t \in \mathcal{T}} \sum_{n=1}^{N_{c,t}} \log\left(\sum_{M \in \mathcal{N}(M^{0}_{c,t,n})} M^{0}_{c,t,n}! \left[\prod_{i \in \mathcal{I}} \frac{\pi_{i}^{M_{i}}}{M_{i}!}\right] \left[\prod_{i \in \mathcal{I}} \exp\left(-\lambda_{c,i,t} \mathcal{D}_{t}\right) \frac{\left(\lambda_{c,i,t} \mathcal{D}_{t}\right)^{M_{i} + M^{1}_{c,i,t,n}}}{(M_{i} + M^{1}_{c,i,t,n})!}\right]\right) \\
& = \ \ -\sum_{c \in \mathcal{C}} \sum_{t \in \mathcal{T}} N_{c,t} S_{c,t} \mathcal{D}_{t} + \sum_{c \in \mathcal{C}} \sum_{t \in \mathcal{T}} \sum_{n=1}^{N_{c,t}} \log\left(\sum_{M \in \mathcal{N}(M^{0}_{c,t,n})} a_{c,t,n}(M) \prod_{i \in \mathcal{I}}  \frac{\left(\lambda_{c,i,t} \mathcal{D}_{t}\right)^{M_{i} + M^{1}_{c,i,t,n}}}{(M_{i} + M^{1}_{c,i,t,n})!}\right) \nonumber
\label{eqn:maxlikelihood5}
\end{align}
where $a_{c,t,n}(M) := M^{0}_{c,t,n}! \prod_{i \in \mathcal{I}} \frac{\pi_{i}^{M_{i}}}{M_{i}!}$.
The maximum likelihood estimator $\hat{\lambda}$ of the model maximizes $\mathscr{L}_{4}$ over the set $\{\lambda \, : \, \lambda \geq 0\}$.
The function $\mathscr{L}_{4}$ is the sum of linear terms and log terms.
To understand the structure of the problem, consider the following simple example with $a_{i} > 0$ and $M_{i,j} \ge 0$:
\begin{align*}
& \max_{\lambda > 0} \left\{-\sum_{j=1}^{n} b_{j} \lambda_{j} + \log\left(\sum_{i=1}^{m} a_{i} \prod_{j=1}^{n} \lambda_{j}^{M_{i,j}}\right)\right\} \\
= \ \ & \max_{\lambda > 0, y > 0} \left\{-\sum_{j=1}^{n} b_{j} \lambda_{j} + \log\left(\sum_{i=1}^{m} a_{i} y_{i}\right) \; : \; y_{i} \le \prod_{j=1}^{n} \lambda_{j}^{M_{i,j}} \; \forall \; i=1,\ldots,m\right\} \\
= \ \ & \max_{\lambda > 0, y > 0} \left\{-\sum_{j=1}^{n} b_{j} \lambda_{j} + \log\left(\sum_{i=1}^{m} a_{i} y_{i}\right) \; : \; \log\left(y_{i}\right) \le \sum_{j=1}^{n} M_{i,j} \log\left(\lambda_{j}\right) \; \forall \; i=1,\ldots,m\right\} \\
= \ \ & \max_{\lambda > 0, \, x} \left\{-\sum_{j=1}^{n} b_{j} \lambda_{j} + \log\left(\sum_{i=1}^{m} a_{i} \exp\left(x_{i}\right)\right) \; : \; x_{i} \le \sum_{j=1}^{n} M_{i,j} \log\left(\lambda_{j}\right) \; \forall \; i=1,\ldots,m\right\}.
\end{align*}
Note that the feasible set $$X := \left\{(\lambda,x) \in \R^{n} \times \R^{m} \; : \; \lambda > 0, \; x_{i} \le \sum_{j=1}^{n} M_{i,j} \log\left(\lambda_{j}\right) \; \forall \; i=1,\ldots,m\right\}$$ of the problem above is convex for any $M_{i,j} \ge 0$.
It is easy to check that a function such as $f(\lambda,x) = -\sum_{j=1}^{n} b_{j} \lambda_{j} + \log\left(\sum_{i=1}^{m} a_{i} \exp\left(x_{i}\right)\right)$ is convex for any weights $a_{i} > 0$.
Therefore the problem of maximizing $\mathscr{L}_{4}$ over the set $\{\lambda : \lambda > 0\}$ can be formulated as the maximization of a convex function over a convex set.
It is well known that the solutions of such a problem are extreme points of the feasible set.
Next we show that corresponding to every $\lambda > 0$ there is an extreme point of $X$, and thus nothing is lost by maximizing over the extreme points of $X$.
Consider any $\lambda > 0$, and let $x \in \R^{m}$ satisfy $x_{i} = \sum_{j=1}^{n} M_{i,j} \log\left(\lambda_{j}\right)$ for all $i=1,\ldots,m$ (that is, all the constraints involving $x$ are active).
Next we show that $(\lambda,x)$ is an extreme point of $X$.
Suppose there exists two points $(\lambda^{1},x^{1}), (\lambda^{2},x^{2}) \in X$ and $\alpha \in (0,1)$ such that $(\lambda^{1},x^{1}) \neq (\lambda^{2},x^{2})$ and $(\lambda,x) = \alpha (\lambda^{1},x^{1}) + (1 - \alpha) (\lambda^{2},x^{2})$.
Then
\begin{align*}
x_{i} \ \ & = \ \ \sum_{j=1}^{n} M_{i,j} \log\left(\lambda_{j}\right)
\ \ = \ \ \sum_{j=1}^{n} M_{i,j} \log\left(\alpha \lambda^{1}_{j} + (1 - \alpha) \lambda^{2}_{j}\right) \\
\ \ & > \ \ \sum_{j=1}^{n} M_{i,j} \left[\alpha \log\left(\lambda^{1}_{j}\right) + (1 - \alpha) \log\left(\lambda^{2}_{j}\right)\right]
\ \ \ge \ \ \alpha x^{1}_{i} + (1 - \alpha) x^{2}_{i}
\ \ = \ \ x_{i}
\end{align*}
which is a contradiction.
Thus, for every $\lambda > 0$, the point $(\lambda,x)$ with $x_{i} = \sum_{j=1}^{n} M_{i,j} \log\left(\lambda_{j}\right)$ for all~$i$ is an extreme point of $X$.

Another challenge with maximizing $\mathscr{L}_{4}$ is computing the sum over $\mathcal{N}(M^{0}_{c,t,n})$, which is a large set if $M_{c,t,n}^{0}$ is large.
To address this challenge, we use Monte Carlo sampling as follows.
Given $\pi := (\pi_{i}, \, i \in \mathcal{I})$ and $\bar{M} \in \mathbb{N}$, let $\mu(\pi,\bar{M})$ denote the multinomial distribution on $\mathcal{N}(M)$ with parameters $\pi$ and $\bar{M}$, that is, for any $M \in \mathcal{N}(\bar{M})$,
\[
\mu(\pi,\bar{M})(M) \ \ = \ \bar{M}! \prod_{i \in \mathcal{I}} \frac{\pi_{i}^{M_{i}}}{M_{i}!}.
\]
For each $c \in \mathcal{C}$, $t \in \mathcal{T}$, and $n = 1,\ldots,N_{c,t}$, let $(M^{s}_{i}, i \in \mathcal{I}), s = 1,\ldots,S_{c,t,n}$ denote a sample of $S_{c,t,n}$ i.i.d.\ observations from $\mu\left(\pi,M^{0}_{c,t,n}\right)$.
Then,
\begin{align}
& \sum_{M \in \mathcal{N}(M^{0}_{c,t,n})} M^{0}_{c,t,n}! \left[\prod_{i \in \mathcal{I}} \frac{\pi_{i}^{M_{i}}}{M_{i}!}\right] \left[\prod_{i \in \mathcal{I}} \exp\left(-\lambda_{c,i,t} \mathcal{D}_{t}\right) \frac{\left(\lambda_{c,i,t} \mathcal{D}_{t}\right)^{M_{i} + M^{1}_{c,i,t,n}}}{(M_{i} + M^{1}_{c,i,t,n})!}\right] \nonumber \\
& = \ \ \mathbb{E}_{M \sim \mu\left(\pi,M^{0}_{c,t,n}\right)} \left[\prod_{i \in \mathcal{I}} \exp\left(-\lambda_{c,i,t} \mathcal{D}_{t}\right) \frac{(\lambda_{c,i,t} \mathcal{D}_{t})^{M_{i} + M^{1}_{c,i,t,n}}}{(M_{i} + M^{1}_{c,i,t,n})!}\right] \nonumber \\
& \approx \ \ u_{c,t,n}(\lambda)
\ \ := \ \ \left[\prod_{i \in \mathcal{I}} \exp\left(-\lambda_{c,i,t} \mathcal{D}_{t}\right)\right] \frac{1}{S_{c,t,n}} \sum_{s = 1}^{S_{c,t,n}} \prod_{i \in \mathcal{I}} \frac{(\lambda_{c,i,t} \mathcal{D}_{t})^{M^{s}_{i} + M^{1}_{c,i,t,n}}}{(M^{s}_{i} + M^{1}_{c,i,t,n})!}
\label{approxuctn}
\end{align}
We implemented a projected gradient method with line search along the feasible direction to solve (find a local minimum of)
\[
\min\left\{\hat{\ell}_{4}(\lambda) \ := \ \sum_{c \in \mathcal{C}} \sum_{t \in \mathcal{T}} N_{c,t} S_{c,t} \mathcal{D}_{t} - \sum_{c \in \mathcal{C}} \sum_{t \in \mathcal{T}} \sum_{n=1}^{N_{c,t}} \log\left(u_{c,t,n}(\lambda)\right) \ : \ \lambda_{c,i,t} \geq \varepsilon\right\}
\]
for $\varepsilon$ small.
The reason for the replacement of the feasible set $\{\lambda : \lambda_{c,i,t} \geq 0\}$ with the feasible set $\{\lambda : \lambda_{c,i,t} \geq \varepsilon\}$ is to ensure that the problem has a differentiable objective on the feasible set so that the projected gradient method can be applied.
Note that
\begin{equation}
\label{derl3u}
\frac{\partial \ell_{4}(\lambda)}{\partial \lambda_{c,i,t}} \ \ = \ \ N_{c,t} \mathcal{D}_{t} - \sum_{n=1}^{N_{c,t}} \frac{1}{u_{c,t,n}(\lambda)} \frac{\partial u_{c,t,n}}{\partial \lambda_{c,i,t}}(\lambda)
\end{equation}
where
\[
\frac{\partial u_{c,t,n}}{\partial \lambda_{c,i,t}}(\lambda) \ \ = \ \ \left[\prod_{j \in \mathcal{I}} \exp\left(-\lambda_{c,j,t} \mathcal{D}_{t}\right)\right] \frac{1}{S_{c,t,n}} \sum_{s = 1}^{S_{c,t,n}} \left(\frac{M^{s}_{i} + M^{1}_{c,i,t,n}}{\lambda_{c,i,t}} - \mathcal{D}_{t}\right) \prod_{j \in \mathcal{I}} \frac{(\lambda_{c,j,t} \mathcal{D}_{t})^{M^{s}_{j} + M^{1}_{c,j,t,n}}}{(M^{s}_{j} + M^{1}_{c,j,t,n})!}.
\]

\if{
$$
u'_{c,t,i,n} = 
\frac{1}{S_{c,t,n}} \sum_{s = 1}^{S_{c,t,n}} f_{c,i,t,n,s}
\prod_{j \in \mathcal{I}}  \frac{\mathcal{D}_{t}^{M^{s}_{j} + M^{1}_{c,j,t,n}}}{(M^{s}_{j} + M^{1}_{c,j,t,n})!}\prod_{j \in \mathcal{I}, j \neq i} e^{-\lambda_{c,j,t} \mathcal{D}_{t}} \lambda_{c,j,t}^{M^{s}_{j} + M^{1}_{c,j,t,n}}
$$
with
$$
f_{c,i,t,n,s} =e^{-\lambda_{c,i,t} \mathcal{D}_{t}} \lambda_{c,i,t}^{M^{s}_{i} + M^{1}_{c,i,t,n}}\Big(-\mathcal{D}_{t} \lambda_{c,i,t}+M_{i}^s+M_{c,i,t,n}^{1}\Big).
$$
}\fi

\if{
Version~2:
\begin{align}
& \frac{\partial \hat{\ell}_{4}}{\partial \lambda_{c,i,t}}(\lambda) \ \ = \ \ N_{c,t} \mathcal{D}_{t} - \sum_{n=1}^{N_{c,t}} \frac{1}{u_{c,t,n}(\lambda)} \frac{\partial u_{c,t,n}}{\partial \lambda_{c,i,t}}(\lambda) \nonumber \\
& \ \ = \ \ N_{c,t} \mathcal{D}_{t} - \sum_{n=1}^{N_{c,t}} \frac{1}{u_{c,t,n}(\lambda)} \Bigg\{-\mathcal{D}_{t} \left[\prod_{j \in \mathcal{I}} \exp\left(-\lambda_{c,j,t} \mathcal{D}_{t}\right)\right] \frac{1}{S_{c,t,n}} \sum_{s = 1}^{S_{c,t,n}} \prod_{j \in \mathcal{I}} \frac{(\lambda_{c,j,t} \mathcal{D}_{t})^{M^{s}_{j} + M^{1}_{c,j,t,n}}}{(M^{s}_{j} + M^{1}_{c,j,t,n})!} \nonumber \\
& \hspace{40mm} {} + \left[\prod_{j \in \mathcal{I}} \exp\left(-\lambda_{c,j,t} \mathcal{D}_{t}\right)\right] \frac{1}{S_{c,t,n}} \sum_{s = 1}^{S_{c,t,n}} \frac{M^{s}_{i} + M^{1}_{c,i,t,n}}{\lambda_{c,i,t}} \prod_{j \in \mathcal{I}} \frac{(\lambda_{c,j,t} \mathcal{D}_{t})^{M^{s}_{j} + M^{1}_{c,j,t,n}}}{(M^{s}_{j} + M^{1}_{c,j,t,n})!}\Bigg\} \nonumber \\
& \ \ = \ \ N_{c,t} \mathcal{D}_{t} - \sum_{n=1}^{N_{c,t}} \frac{1}{u_{c,t,n}(\lambda)} \left[\prod_{j \in \mathcal{I}} \exp\left(-\lambda_{c,j,t} \mathcal{D}_{t}\right)\right] \frac{1}{S_{c,t,n}} \sum_{s = 1}^{S_{c,t,n}} \left(\frac{M^{s}_{i} + M^{1}_{c,i,t,n}}{\lambda_{c,i,t}} - \mathcal{D}_{t}\right) \prod_{j \in \mathcal{I}} \frac{(\lambda_{c,j,t} \mathcal{D}_{t})^{M^{s}_{j} + M^{1}_{c,j,t,n}}}{(M^{s}_{j} + M^{1}_{c,j,t,n})!}
\label{derl3u}
\end{align}

}\fi

\if{
\subsection*{Função de Verossimilhança Logarítmica Total}

Dividi como sendo a Soma dos processos das localizações reportadas e não reportadas:  

\[
\log \mathscr{L}(\lambda, p) = \sum_{c} \sum_{i} \sum_{t} \sum_{n} \left[ M_{c, i, t, n}^{(1)} \log \left( (1 - p_{c, i}) \lambda_{c, i, t} \right) - (1 - p_{c, i}) \lambda_{c, i, t} \right]
\]
\[
+ \sum_{c} \sum_{i} \sum_{t} \sum_{n} \left[ M_{c, t, n}^{(0)} \log \left( p_{c, i} \lambda_{c, i, t} \right) - p_{c, i} \lambda_{c, i, t} \right]
\]
Deduzi a partir do produtório de cada Verossimilhança com um processo de poisson.

\subsection*{Hipóteses: }
Distribuição "real", onde os dados são originados:
  \[
  Y_{k} \sim \text{Poisson}(\lambda_{k}), \quad \text{para } k = 1, 2, \dots, K.
  \]
"Distribuição Consequência"
  \[
  X_{k} \mid Y_{k} \sim \text{Binomial}(Y_{k}, 1 - p_{k}), \quad \text{para } k = 1, 2, \dots, K.
  \]

\subsection*{EM}

- Inicialize os parâmetros \( \lambda_{k}^{(0)} \) e \( p_{k}^{(0)} \)- pode ser escolhido de forma "esperta"

ETAPA E: 
 A expectativa condicional é:
  \[
  \mathbb{E}[Y_{k} \mid X_{k}] = \frac{X_{k}}{1 - p_{k}^{(t)}}, \quad \text{para } k = 1, 2, \dots, K.
  \]
Ou seja, se foi observado um valor em X, então provavelmente Esse valor real é muito maior

Etapa M: 
-Iteração: 
  \[
  \lambda_{k}^{(t+1)} = \frac{1}{n} \sum_{i=1}^{n} \mathbb{E}[Y_{k,i} \mid X_{k,i}] = \frac{1}{n} \sum_{i=1}^{n} \frac{X_{k,i}}{1 - p_{k}^{(t)}}, \quad \text{para } k = 1, 2, \dots, K.
  \]

Eu comecei tentando fazer isso:(mas acabei mudando de ideia)
  \[
  p_{k}^{(t+1)} = 1 - \frac{1}{n \cdot \lambda_{k}^{(t+1)}} \sum_{i=1}^{n} X_{k,i}, \quad \text{para } k = 1, 2, \dots, K.
  \]

Isso foram as estimativas teóricas, mas eu acabei implementando a seguinte iteração no meu código: 
Iteração sobre \( \lambda_{c,i,t} \):
   \[
   \lambda_{c,i,t}^{(t+1)} = (1 - \text{learning\_rate}) \cdot \lambda_{c,i,t}^{(t)} + \text{learning\_rate} \cdot \left( \frac{\sum_n M_{c,i,t,1}}{n_{\text{observations}} \cdot (1 - p_{c,i}^{(t)})} \right)
   \]

Iteração sobre \( p_{c,i} \)
   \[
   p_{c,i}^{(t+1)} = (1 - \text{learning\_rate}) \cdot p_{c,i}^{(t)} + \text{learning\_rate} \cdot \left( \frac{\sum_{t} M_{c,t,0}[c,i,:]}{\sum_{t} M_{c,t,0}[c,i,:] + \sum_{t} M_{c,i,t,1}[c,i,:] + \epsilon} \right)
   \]
(isso é similar à solução anaítica com um único p, pois estamos maximizando a verossimihança, fixando os lambdas )

Aqui, estou tentando atualizar os parâmetros por uma combinação convexa das iterações, para ir mais lentamente em direção ao objetivo

O critério de parada é a estabilização da Log- verossimilhança, por um hiperparâmetro $\epsilon$

\section*{Ideias: }
E se considerarmos uma mixture do tipo: 
Probabilidade 1-p de reportar, e se reportar ele segue uma Poisson? 

1. O vetor de parâmetros é \( \theta = (p, \mu) \).

2. Omitindo o coeficiente multinomial óbvio, a log-verossimilhança para os dados observados é:
   \[
   \ln g(y \mid \theta) = y_{0} \ln \left[p + (1 - p) e^{-\mu}\right] + \sum_{k \geq 1} y_{k} \left[\ln (1 - p) + k \ln \mu - \mu - \ln k!\right]
   \]

3. Não existe uma forma fechada para o máximo dessa log-verossimilhança.

 Os Dados Completos - ---------------------------------

1. Para gerar os dados completos, dividimos os \( y_{0} \) indivíduos entre \( x_{A} \) indivíduos da população \( A \) e \( x_{B} \) indivíduos da população \( B \).

2. A log-verossimilhança para os dados completos é:
   \[
   \ln f(x \mid \theta) = x_{A} \ln p + x_{B} \left[\ln (1 - p) - \mu\right] + \sum_{k \geq 1} y_{k} \left[\ln (1 - p) + k \ln \mu - \mu - \ln k!\right]
   \]

3. Na etapa E, calculamos:
   \[
   x_{A}^{n} = \mathbb{E}\left(x_{A} \mid y_{0}, \theta^{n}\right) = y_{0} \frac{p^{n}}{p^{n} + \left(1 - p^{n}\right) e^{-\mu^{n}}}
   \]
   \[
   x_{B}^{n} = \mathbb{E}\left(x_{B} \mid y_{0}, \theta^{n}\right) = y_{0} - \mathbb{E}\left(x_{A} \mid y_{0}, \theta^{n}\right)
   \]

 A Etapa M-----------------------------

1. A função substituta (surrogate function) é:
   \[
   Q\left(\theta \mid \theta^{n}\right) = x_{A}^{n} \ln p + x_{B}^{n} \left[\ln (1 - p) - \mu\right] + \sum_{k \geq 1} y_{k} \left[\ln (1 - p) + k \ln \mu - \mu - \ln k!\right]
   \]

2. O máximo ocorre para:
   \[
   p^{n+1} = \frac{x_{A}^{n}}{y_{0} + \sum_{k \geq 1} y_{k}}
   \]
   \[
   \mu^{n+1} = \frac{\sum_{k \geq 1} k y_{k}}{x_{B}^{n} + \sum_{k \geq 1} y_{k}}
   \]

 Estimativa de Parâmetros com p "conhecido" ------------------------

Queremos encontrar estimadores estáveis para os parâmetros \(\lambda = (\lambda_{j})_{j}\) no seguinte problema. Consideramos uma matriz \(A\) de pesos, onde cada peso \(a_{i j}\) é proporcional à área da zona \(i\). Especificamente, definimos:

\[
a_{i j} = \frac{A_{i}}{A}
\]

onde \(A_{i}\) é a área da zona \(i\) e \(A\) é a soma das áreas de todas as zonas. Os \(N_{i j}\) seguem a distribuição de Poisson:

\[
N_{i j} \sim \mathcal{P}\left(a_{i j} \lambda_{j}\right) = \mathcal{P}\left(\frac{A_{i}}{A} \lambda_{j}\right)
\]

Como não conhecemos as observações \(N_{i j}\), temos as observações \( \{Y_{i}\}_{i=1, \ldots, n} \) com a seguinte distribuição:

\[
Y_{i} = \sum_{j=1}^{m} N_{i j} \sim \mathcal{P}\left(\sum_{j=1}^{m} a_{i j} \lambda_{j}\right) = \mathcal{P}\left(\sum_{j=1}^{m} \frac{A_{i}}{A} \lambda_{j}\right)
\]

Queremos encontrar estimadores estáveis para \(\lambda\) usando os dados \(\{Y_{i}\}_{i=1, \ldots, n}\).

Função de Verossimilhança-----------------------------

Assuma: 

\[
\mathscr{L}_{\left(N_{i j}\right)_{i j}}(\lambda) = \prod_{i=1}^{n} \prod_{j=1}^{m} e^{\frac{A_{i}}{A} \lambda_{j}} \frac{\left(\frac{A_{i}}{A} \lambda_{j}\right)^{N_{i j}}}{N_{i j}!}
\]

Logo: 

\[
l_{\left(N_{i j}\right)_{i j}}(\lambda) = \sum_{i=1}^{n} \sum_{j=1}^{m} \left(-\frac{A_{i}}{A} \lambda_{j} + N_{i j} \log \left(\frac{A_{i}}{A} \lambda_{j}\right) - \log \left(N_{i j}!\right)\right)
\]

*otimizando:* -----------------------------------------

A derivada da log-verossimilhança com respeito a \(\lambda_{j}\) é:

\[
\frac{d}{d \lambda_{j}} E\left[l_{\left(N_{i j}\right)_{i j}} \mid \left(Y_{i}\right)_{i}\right] = \sum_{i=1}^{n} -\frac{A_{i}}{A} + \frac{1}{\lambda_{j}} E\left[N_{i j} \mid \left(Y_{i}\right)_{i}\right]
\]

Usando a distribuição condicional \(N_{i j} \mid Y_{i}\), que segue uma distribuição binomial:

\[
N_{i j} \mid Y_{i} \sim \operatorname{Bin}\left(Y_{i}, \frac{a_{i j} \lambda_{j}}{\sum_{k=1}^{m} a_{i k} \lambda_{k}}\right)
\]

Temos a expectativa:

\[
E\left[N_{i j} \mid Y_{i}\right] = \frac{Y_{i} \frac{A_{i}}{A} \lambda_{j}}{\sum_{k=1}^{m} \frac{A_{i}}{A} \lambda_{k}}
\]

Substituindo isso na derivada da log-verossimilhança:

\[
\frac{d}{d \lambda_{j}} E\left[l_{\left(N_{i j}\right)_{i j}} \mid \left(Y_{i}\right)_{i}\right] = \sum_{i=1}^{n} -\frac{A_{i}}{A} + \frac{1}{\lambda_{j}} \frac{Y_{i} \frac{A_{i}}{A} \lambda_{j}}{\sum_{k=1}^{m} \frac{A_{i}}{A} \lambda_{k}}
\]

Método iterativo: -------------------------------------------

Definindo a atualização dos parâmetros \(\lambda_{j}\) resolvendo a equação para a derivada igual a zero:

\[
0 = \sum_{i=1}^{n} -\frac{A_{i}}{A} + \frac{1}{\lambda_{j}} \frac{Y_{i} \frac{A_{i}}{A} \lambda_{j}}{\sum_{k=1}^{m} \frac{A_{i}}{A} \lambda_{k}}
\]

Resolvendo para \(\lambda_{j}\):

\[
\lambda_{j} = \frac{\lambda_{j}^{\text{old}}}{\sum_{i=1}^{n} \frac{A_{i}}{A}} \sum_{i=1}^{n} \frac{Y_{i} \frac{A_{i}}{A}}{\sum_{k=1}^{m} \frac{A_{i}}{A} \lambda_{k}^{\text{old}}}
\]

Portanto, incorporamos a notação dos pesos proporcionais à área na formulação original e apresentamos a função de verossimilhança correspondente.

--------------------------------------

\begin{itemize}
    \item Latent Variables: \( X_{c, i, t, n} \sim \text{Poisson}(p \cdot \lambda_{c, i, t} \mathcal{D}_{t}) \), the number of arrivals for type \( c \), in zone \( i \), in time interval \( t \), and in observation \( n \), that are not reported.
    \item Observable Variables:
    \begin{itemize}
        \item \( Y_{c, i, t, n} \sim \text{Poisson}((1 - p) \cdot \lambda_{c, i, t} \mathcal{D}_{t}) \), the number of arrivals for type \( c \), in zone \( i \), in time interval \( t \), and in observation \( n \), that are reported.
        \item \( Z_{c, t, n} \sim \text{Poisson}(p \cdot S_{c, t} \cdot \mathcal{D}_{t}) \), the number of arrivals for type \( c \), in time interval \( t \), and in observation \( n \), that are not reported, summing over all zones.
    \end{itemize}
\end{itemize}

 \( S_{c, t} = \sum_{i \in \mathcal{I}} \lambda_{c, i, t} \) is the sum of the Poisson process intensities across all zones for type \( c \) and time interval \( t \).

\textbf{E-Step (Expectation)}
In the \textbf{E-Step}, the goal is to calculate the conditional expectation of the latent variables \( X_{c, i, t, n} \), given the total observed \( Z_{c, t, n} \) and the current parameters \( \lambda_{c, i, t} \).

For each type \( c \), zone \( i \), time interval \( t \), and observation \( n \):

Conditional Expectation of \( X_{c, i, t, n} \):
   The conditional expectation of the latent variables \( X_{c, i, t, n} \), given \( Z_{c, t, n} \) and parameters \( \lambda_{c, i, t} \), is:

   \[
   \mathbb{E}[X_{c, i, t, n} \mid Z_{c, t, n}] = Z_{c, t, n} \cdot \frac{p \cdot \lambda_{c, i, t}}{S_{c, t}}.
   \]

   This represents the expected number of unreported arrivals for type \( c \), in zone \( i \), in time interval \( t \), and in observation \( n \). With this, we update the sufficient statistics for each \( \lambda_{c, i, t} \), which will be utilized in the \textbf{M-Step} to adjust the parameters.

\textbf{M-Step (Maximization)}

Coming from the Likelihood function,  $ \log P(X_{c, i, t, n}, Y_{c, i, t, n} \mid \lambda_{c, i, t}) $,  Summing over all observations and zones, the complete log-likelihood function becomes:

   \[
   Q(\lambda_{c, i, t}) = \sum_{c \in \mathcal{C}} \sum_{t \in \mathcal{T}} \sum_{i \in \mathcal{I}} \sum_{n=1}^{N_{c, t}} \left( \mathbb{E}[X_{c, i, t, n}] \log(p \cdot \lambda_{c, i, t} \cdot \mathcal{D}_{t}) + Y_{c, i, t, n} \log((1 - p) \cdot \lambda_{c, i, t} \cdot \mathcal{D}_{t}) \right).
   \]

   To maximize \( Q(\lambda_{c, i, t}) \), we obtain

   \[
   \lambda_{c, i, t} = \sum_{n=1}^{N_{c, t}} \left(\mathbb{E}[X_{c, i, t, n}] + Y_{c, i, t, n}\right).
   \]

\textbf{{EM Algorithm Steps}}

1.\textbf{Initialization:} Choose initial values for the parameters \( \lambda_{c, i, t} \).
\newline 
2. \textbf{E-Step:}
   \begin{itemize}
       \item Compute \( \mathbb{E}[X_{c, i, t, n} \mid Z_{c, t, n}] = Z_{c, t, n} \cdot \frac{p \cdot \lambda_{c, i, t}}{S_{c, t}} \).
       \item Update the sufficient statistics \( \mathbb{E}[X_{c, i, t, n}] \).
   \end{itemize}
3. \textbf{M-Step:}
   \begin{itemize}
       \item Update the parameters \( \lambda_{c, i, t} \) using the formula \( \lambda_{c, i, t} = \sum_{n=1}^{N_{c, t}} \left(\mathbb{E}[X_{c, i, t, n}] + Y_{c, i, t, n}\right) \).
   \end{itemize}
4. \textbf{Iteration:} Repeat the E and M steps until convergence is achieved.

}\fi

\section{Numerical Results: Models for Emergency Call Arrivals}
\label{sec:num}

We used data about medical emergencies reported to the Rio de Janeiro emergency medical service to estimate the models of Section~\ref{sec:model1ml}.
The data include the date and time of the emergency call, with the date ranging from 2016/01/01 to 2018/01/08, the location of the emergency when reported, and the type of the emergency.
The emergency type data includes only a ``priority level'': high, intermediate, and low-priority emergencies.
The calibrated models were periodic with a period of one week.
Time during the week was discretized into time intervals of length $30$~minutes (thus the model had $T = 7 \times 48 = 336$ time intervals).
Our software LASPATED \citep{laspatedpaper,laspatedmanual} facilitates space discretizations into rectangles, hexagons, by districts, or by the Voronoi diagram associated with ambulance
stations (the locations of these ambulance stations is provided in our GitHub repository and on our project website, see \citealt{websiteambrouting24}).
Figure~\ref{figurerect1010} shows a $10 \times 10$ rectangular space discretization containing the city of Rio de Janeiro (76 of these rectangles have a nonempty intersection with the city and are shown in the figure).

\begin{figure}
\centering
\begin{tabular}{c}
\includegraphics[scale=0.8]{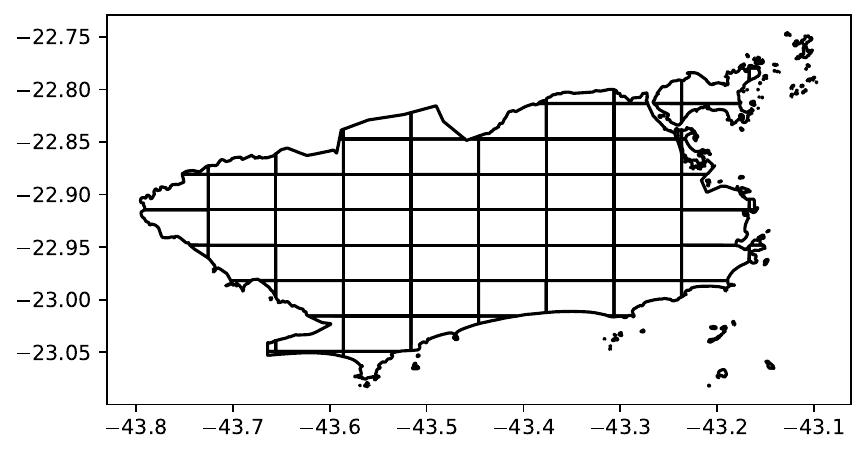}
\end{tabular}
\caption{\label{figurerect1010}
Space discretization of a region containing the city of Rio de Janeiro into $10 \times 10 = 100$ rectangles, 76 of which have nonempty intersection with the region and are shown in the figure.}
\end{figure}

Figure~\ref{empprobmodel1} shows $\hat{p}_{c,t}$ as a function of $t$, for the three values of the arrival type~$c$.
The probabilities $\hat{p}_{c,t}$ of the location not being reported exhibit regular cycles with period $1$~day, oscillating between approximately $0.1$ and $0.5$.
In Figure~\ref{empprobmodel1}, time~$0$ represents Sunday/Monday at midnight, and thus the probabilities of the location not being reported are the largest at night.
Also, there are significant differences between the probabilities $\hat{p}_{c,t}$ for different types~$c$.
The probabilities $\hat{p}_{c,t}$ of the location not being reported are similar for high and intermediate priority emergencies, and significantly larger for low priority emergencies.
Figure~\ref{empprobmodel1} also shows the estimate $\hat{p}$ of the single probability of the location not being reported, given by \eqref{formp}.
Thus, the model with parameters $p_{c,t}$ fit the data better than the model with a single parameter~$p$.

\begin{figure}
    \centering
    \begin{tabular}{c}
        \includegraphics[scale=0.9]{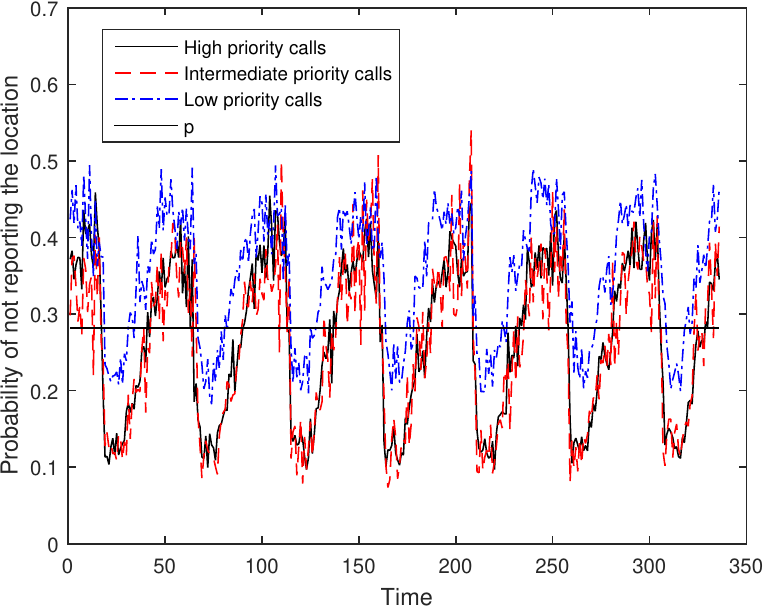}        
     \end{tabular}
 \caption{Estimated probabilities of the emergency location not being reported for high priority, intermediate priority, and low priority emergencies.
Also shown is the estimate of the single probability $p$ of the location not being reported, independent of the time and type of emergency.}
\label{empprobmodel1}
\end{figure}

Next we show the regularized estimators of $\lambda_{c,i,t}$ as described in Section~\ref{sec:reg}, for different values of weights $w_{i,j}$ and $W_{G}$.
Specifically, given weight $w > 0$, for space regularization we chose $w_{i,j} = w$ when zones~$i$ and~$j$ are neighbors, and $w_{i,j} = 0$ otherwise, and for time regularization we chose $W_{G} = w$ for all time groups $G$.
We used the following eight time groups during the week:
\begin{itemize}
\item Time group 1: Monday, Tuesday, Wednesday, Thursday, Friday for the time period 06:00--10:00;
\item Time group 2: Monday, Tuesday, Wednesday, Thursday, Friday for the time period 10:00--18:00;
\item Time group 3: Monday, Tuesday, Wednesday, Thursday, Friday for the time period 18:00--22:00;
\item Time group 4: Monday, Tuesday, Wednesday, Thursday for the time period 22:00--24:00 and 00:00-06:00; Friday for the time period 00:00--06:00; and Sunday for the time period 22:00--24:00;
\item Time group 5: Friday, Saturday for the time period 22:00--24:00, Saturday for the period 
00:00-06:00, and Sunday for the period 
00:00-06:00; 
\item Time group 6: Saturday, Sunday for the time period 06:00--10:00;
\item Time group 7: Saturday, Sunday for the time period 10:00--18:00;
\item Time group 8: Saturday, Sunday for the time period 18:00--22:00.
\end{itemize}
To solve \eqref{pboptinit1}, we used the projected gradient method with Armijo line search along the feasible direction, see for instance \cite{laspatedpaper,laspatedmanual,iusem2003}.
Recall that $\mathscr{L}_{2}$ is defined if $\lambda_{c,i,t} > 0$ and $0 < \hat{p}_{c,t} < 1$.
We chose $\varepsilon > 0$ small and imposed the constraints $\lambda \geq \varepsilon {\textbf{e}}$ and $\varepsilon \leq p_{c,t} \leq 1 - \varepsilon$ on \eqref{pboptinit1}, where \textbf{e} is a vector of ones.
That is, we estimated $\lambda_{c,i,t}$ by solving
\begin{equation}
\label{pboptinitr}
\min_{\lambda \geq \varepsilon {\textbf{e}}} \; \ell_{1}(\lambda)
\end{equation}
and we estimated $p_{c,t}$ by solving
\begin{equation}
\label{pbestp}
\min_{\varepsilon \leq p_{c,t} \leq 1 - \varepsilon} \; \ell_{2}(p).
\end{equation}

Figure \ref{regularized_model} shows the estimated Poisson intensities over the entire region for $w \in \{0$, $0.001$, $0.005$, $0.01$, $0.03\}$.
(Note that $w = 0$ gives the intensities of the unregularized model of Section~\ref{sec:model1ml}.)
When $w$ increases, the Poisson intensities decrease and tend to be closer to each other.

\begin{figure}[hbtp]
    \centering
    \begin{tabular}{cc}
        \includegraphics[scale=0.60]{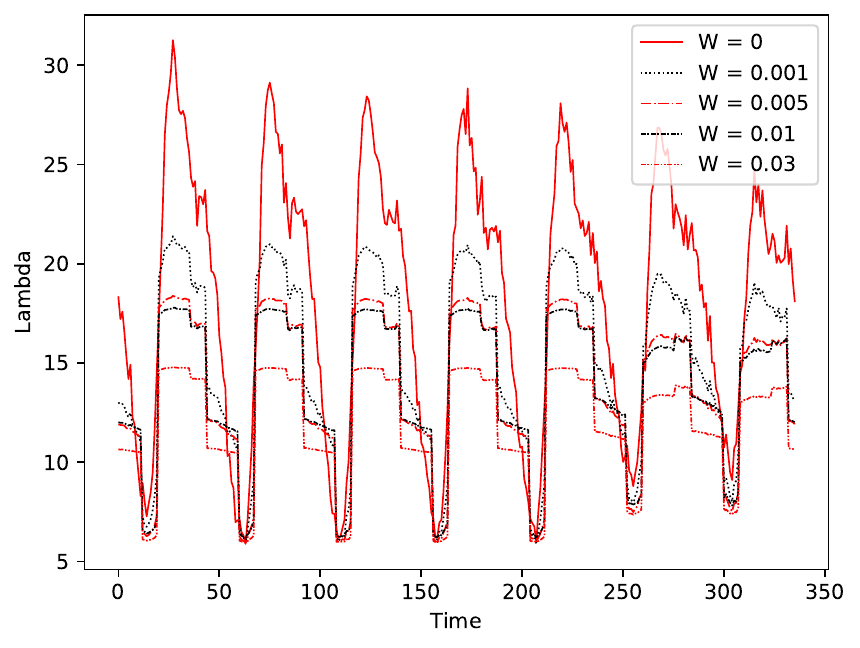} & \includegraphics[scale=0.60]{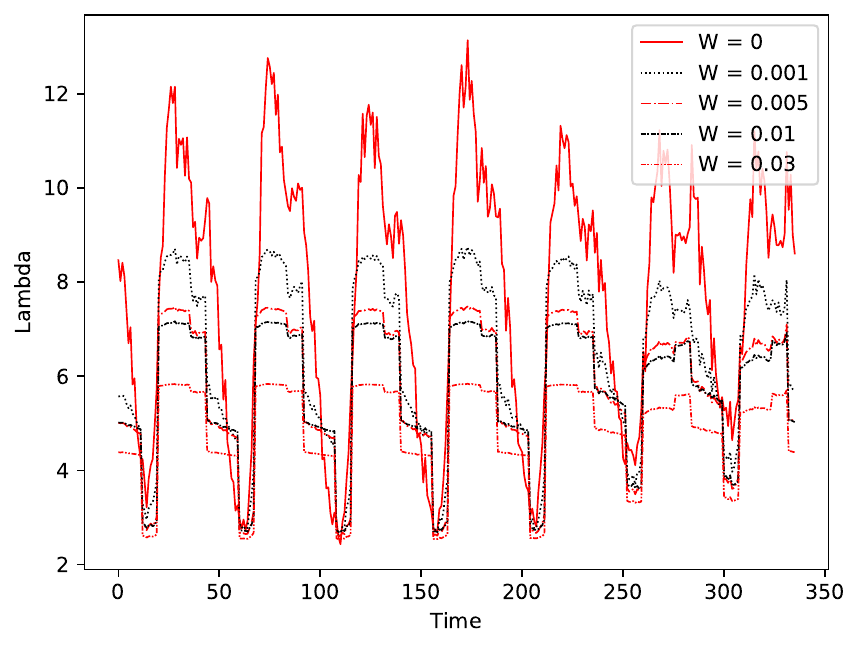}\\     
        \includegraphics[scale=0.60]{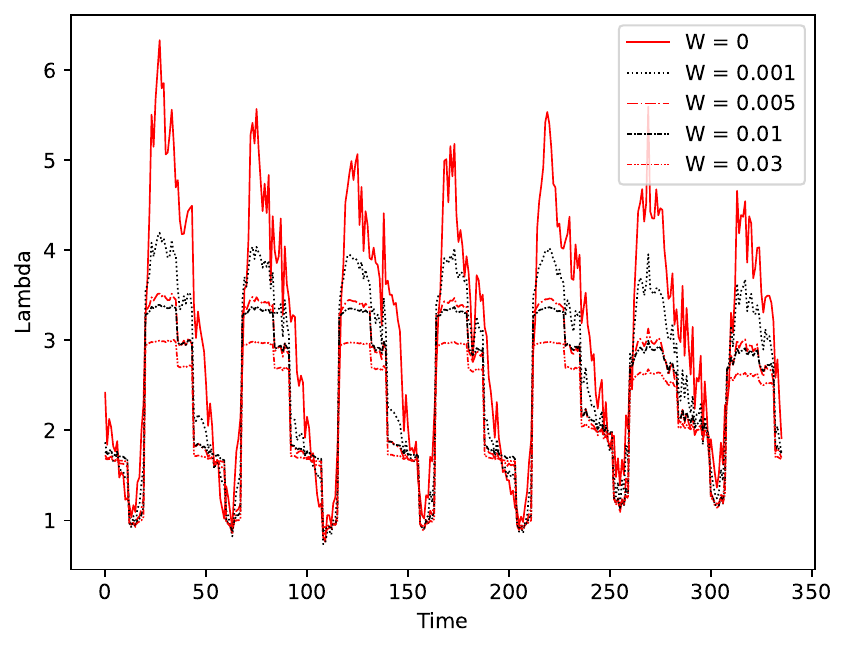} & \includegraphics[scale=0.60]{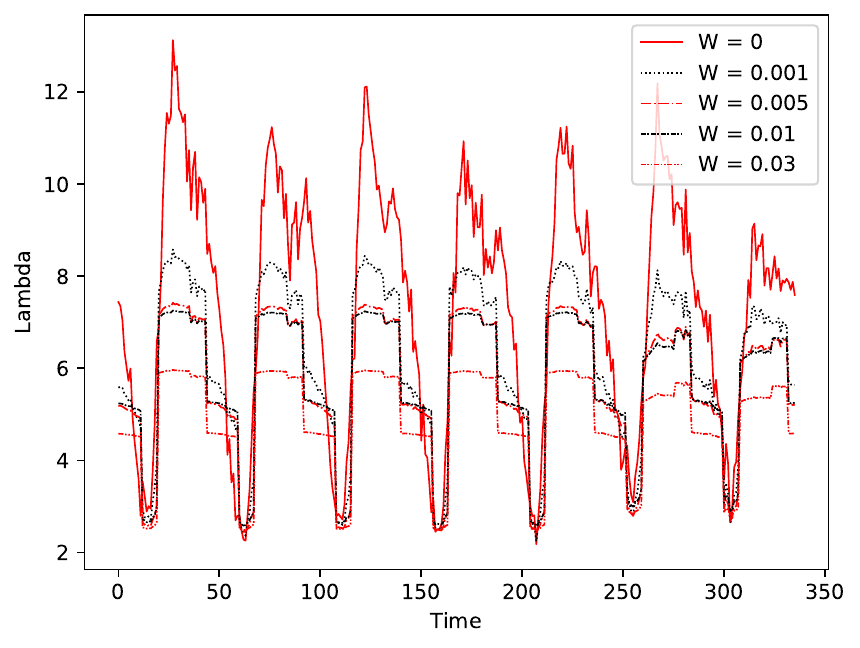}
     \end{tabular}
\caption{Poisson intensities for the regularized model with different weight values $w$. The top-left figure is the sum of intensities over all priorities, the top-right shows the total intensities for high-priority emergencies, the bottom-left shows the total intensities for intermediate-priority emergencies, and the bottom-right plot shows the total intensities for low-priority emergencies.}
\label{regularized_model}
\end{figure}

We also estimated the models of Sections~\ref{sec:modelcov} and~\ref{sec:model1m2}.
These models produce solutions very close to the solution for the model in Section~\ref{sec:model1ml}.

Next, we compare the estimated ``corrected'' intensities for the model in Section~\ref{sec:model1ml} with the estimated ``uncorrected'' intensities obtained after discarding all arrivals with no location reported.
Figure~\ref{emplammodel1} shows the 95\% confidence intervals (the 0.05 and 0.95 quantiles) for the Poisson intensities over the entire region as a function of 30 minute time intervals during the week.
These results are reported for all emergencies, high priority emergencies, intermediate priority emergencies, and low priority emergencies, for the corrected and uncorrected estimates.
As expected, the corrected estimates are significantly higher than the uncorrected estimates.

\begin{figure}
    \centering
    \begin{tabular}{cc}
        \includegraphics[scale=0.65]{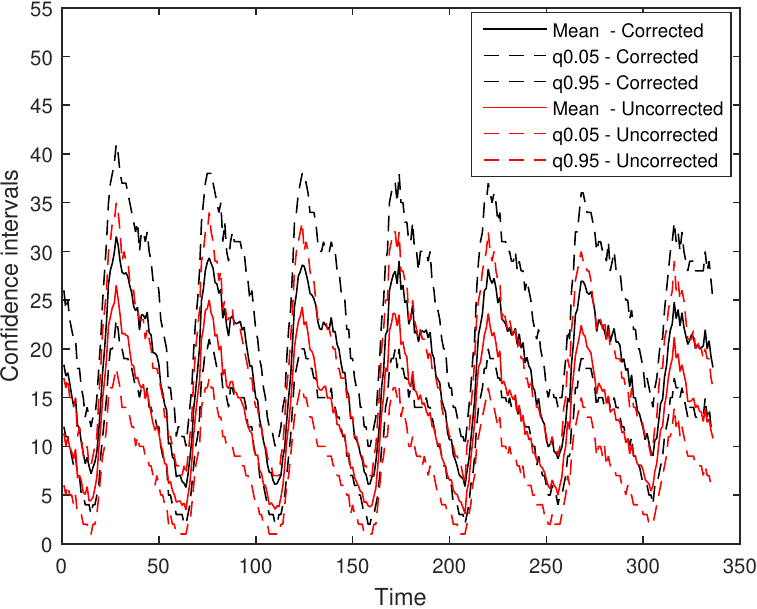} & \includegraphics[scale=0.65]{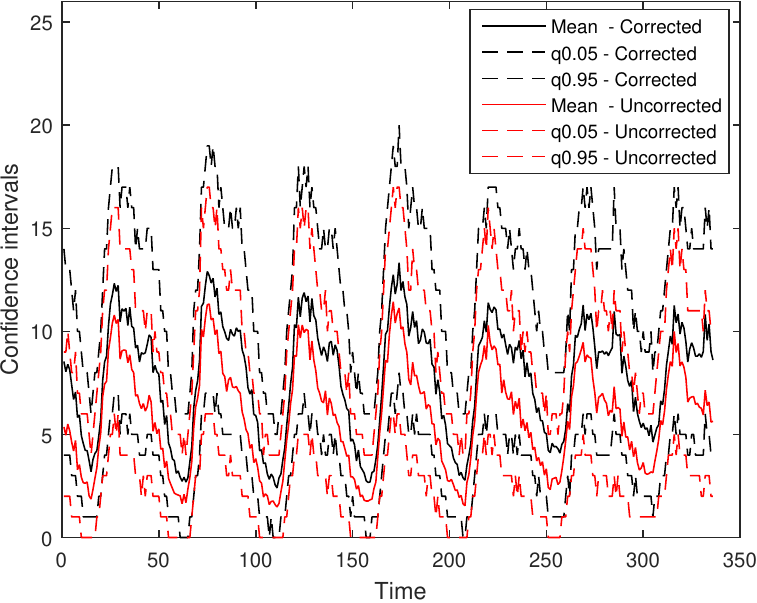}\\     
        \includegraphics[scale=0.65]{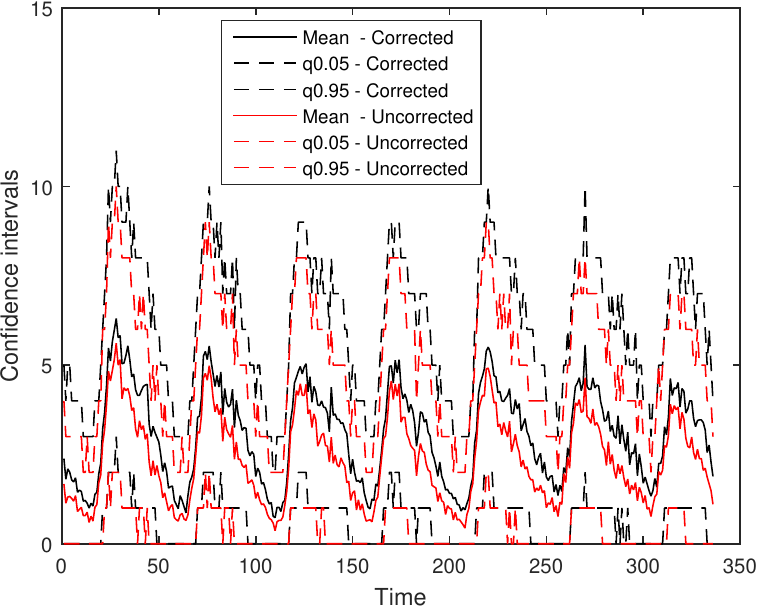} & \includegraphics[scale=0.65]{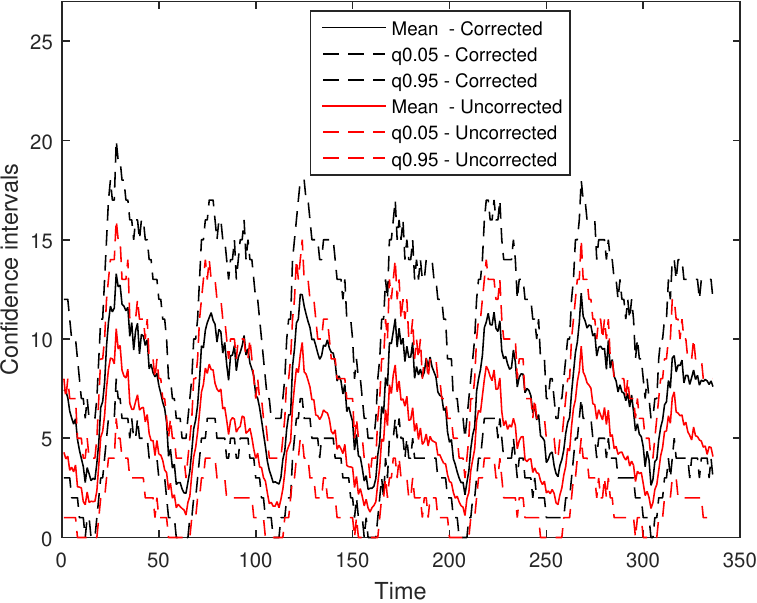}
     \end{tabular}
\caption{Confidence intervals for the arrival intensities over the entire region as a function of the 30 minute time intervals during the week.
Top left: all emergencies.
Top right: high priority emergencies.
Bottom left: intermediate priority emergencies.
Bottom right: low priority emergencies.}
\label{emplammodel1}
\end{figure}

Figure~\ref{histmodel1} shows histograms for the corrected and uncorrected estimates of total numbers of arrivals over the entire region and the entire week, for all emergencies and for the three types of emergencies.
Since the data contain a large number of emergencies with no location reported, the corrected histograms are significantly shifted to the right relative to the uncorrected histograms.

\begin{figure}
    \centering
    \begin{tabular}{cc}
        \includegraphics[scale=0.63]{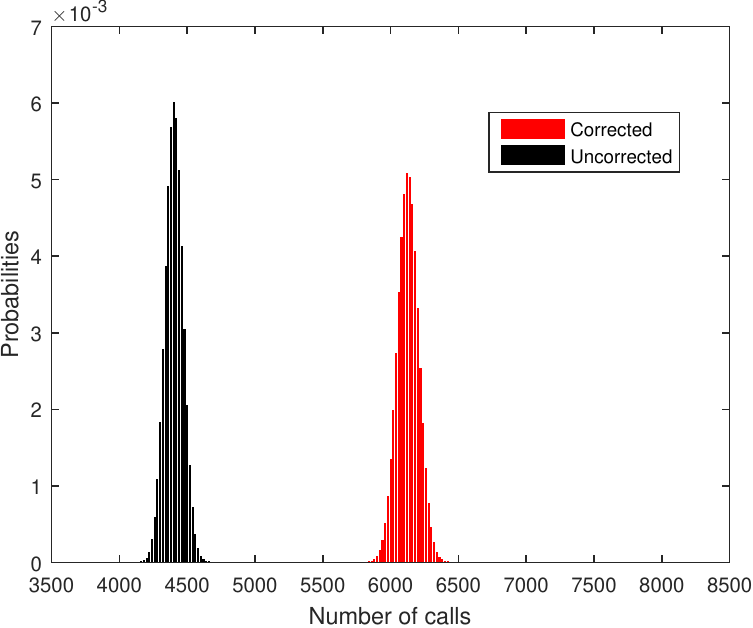} & \includegraphics[scale=0.63]{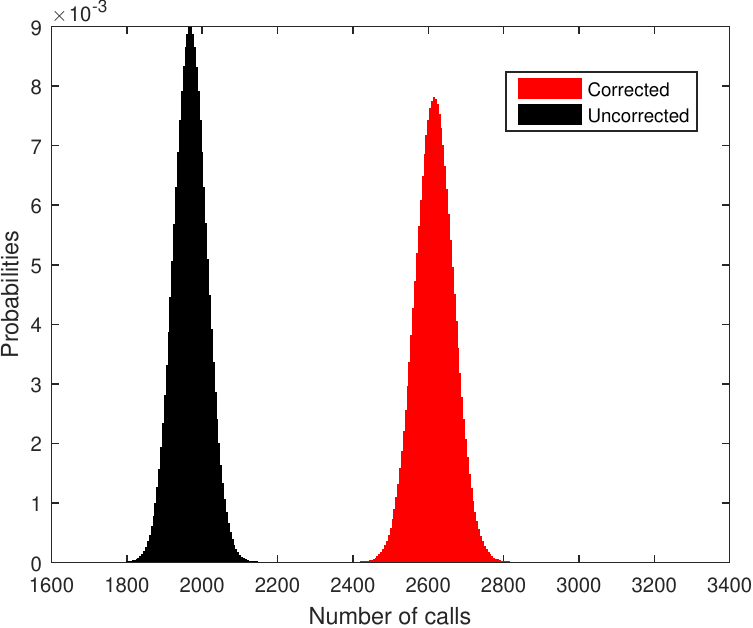}\\     
        \includegraphics[scale=0.63]{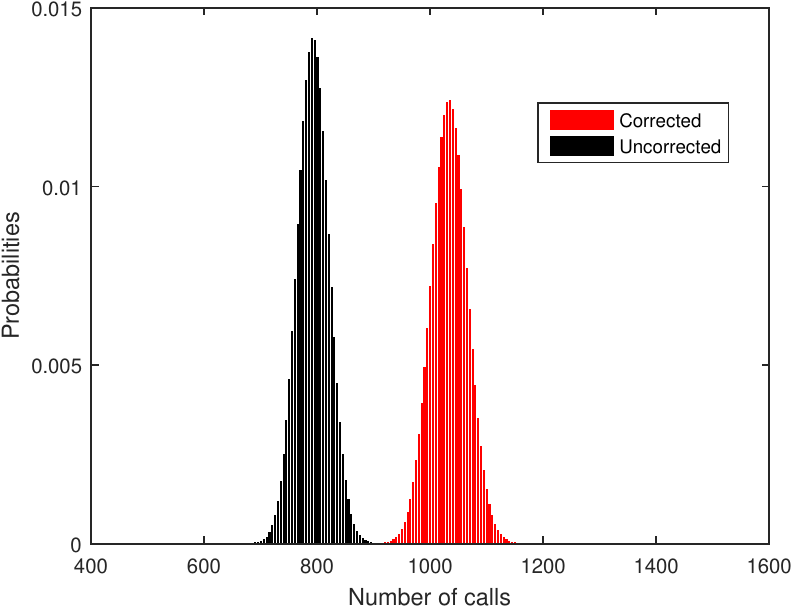} & \includegraphics[scale=0.63]{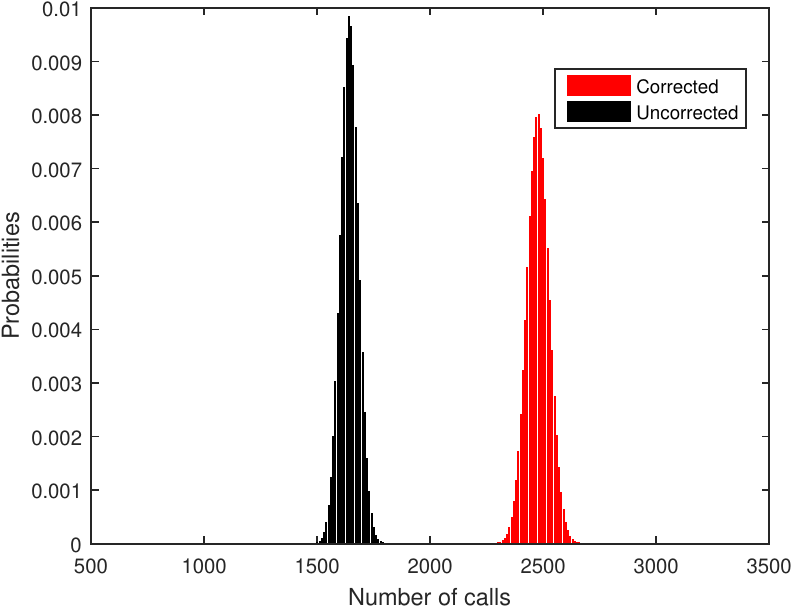}
     \end{tabular}
\caption{Histograms of the total number of emergencies over the entire region and the entire week using data with missing locations (in red) and without using these data (in black).
Top left: all emergencies.
Top right: high priority emergencies.
Bottom left: intermediate priority emergencies.
Bottom right: low priority emergencies.}
\label{histmodel1}
\end{figure}

Figures~\ref{heat1model0} and~\ref{heat1model1} show heatmaps of the corrected and uncorrected intensity estimates for all emergencies, and for the three types of emergencies.
The legends on the right of the figures indicate that the corrected intensity estimates are significantly higher than the uncorrected intensity estimates, while the space distribution is similar for both types of estimates.

\begin{figure}
    \centering
    \begin{tabular}{cc}
        \includegraphics[scale=0.3]{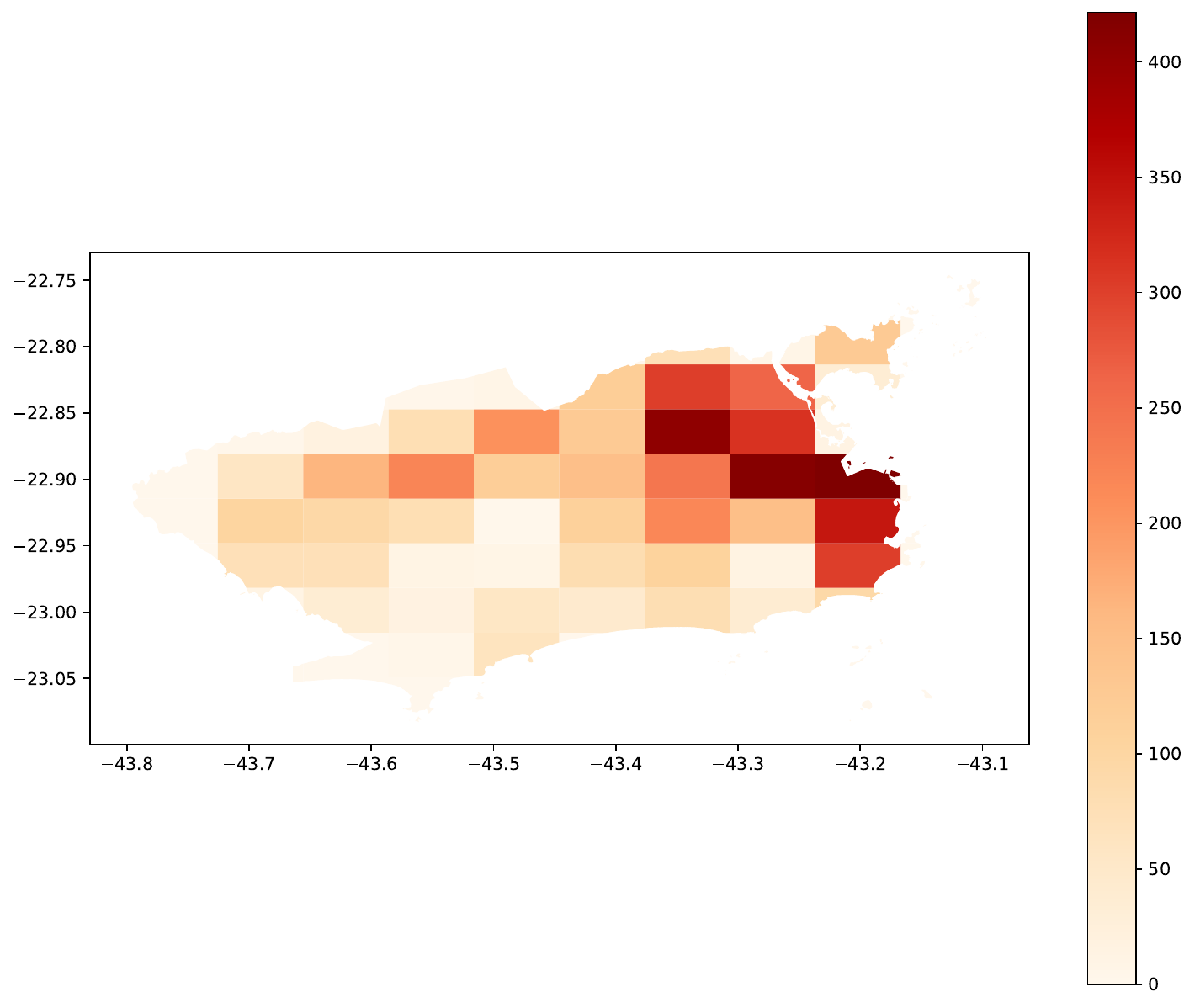} & \includegraphics[scale=0.3]{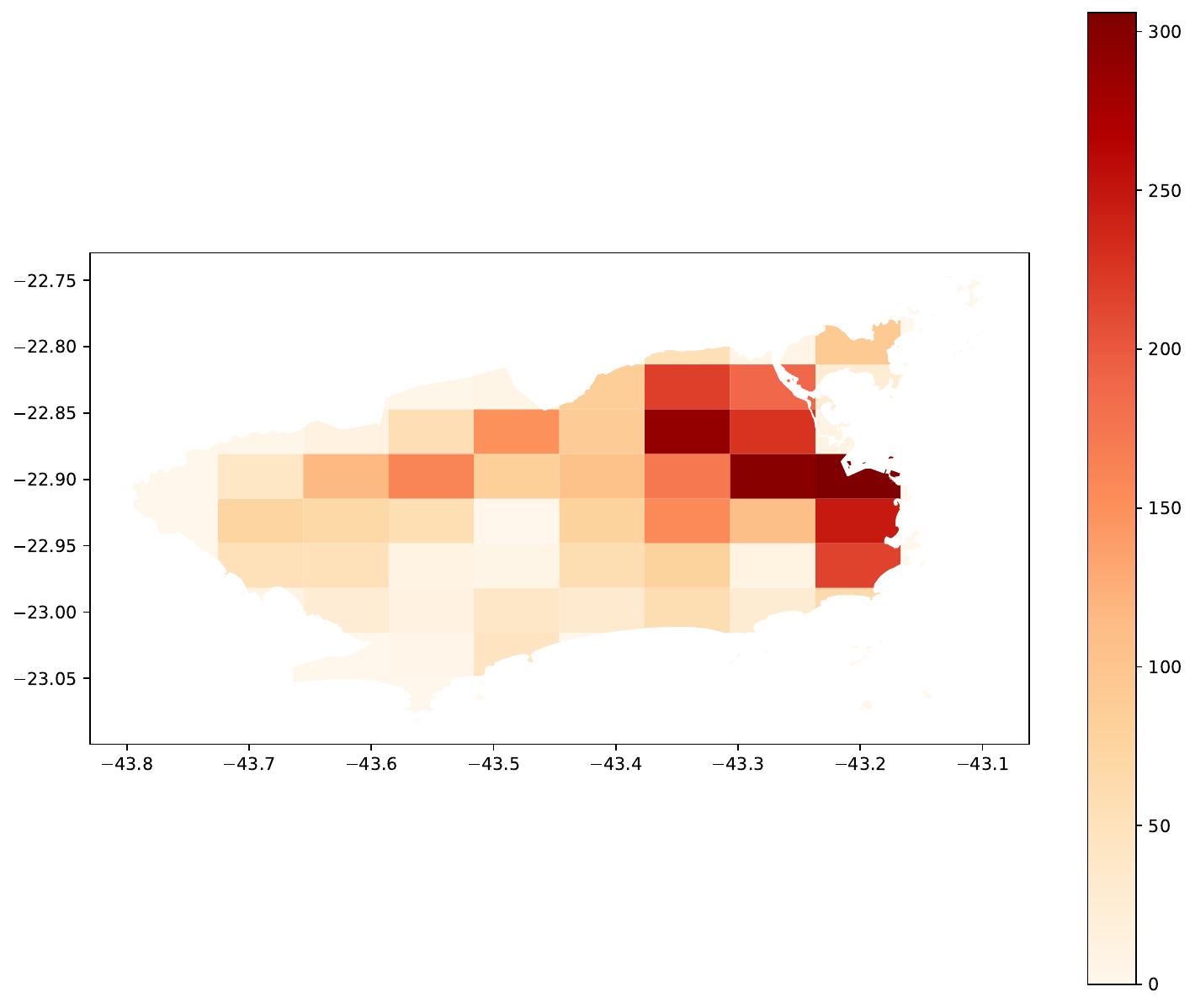}\\     
    \includegraphics[scale=0.3]{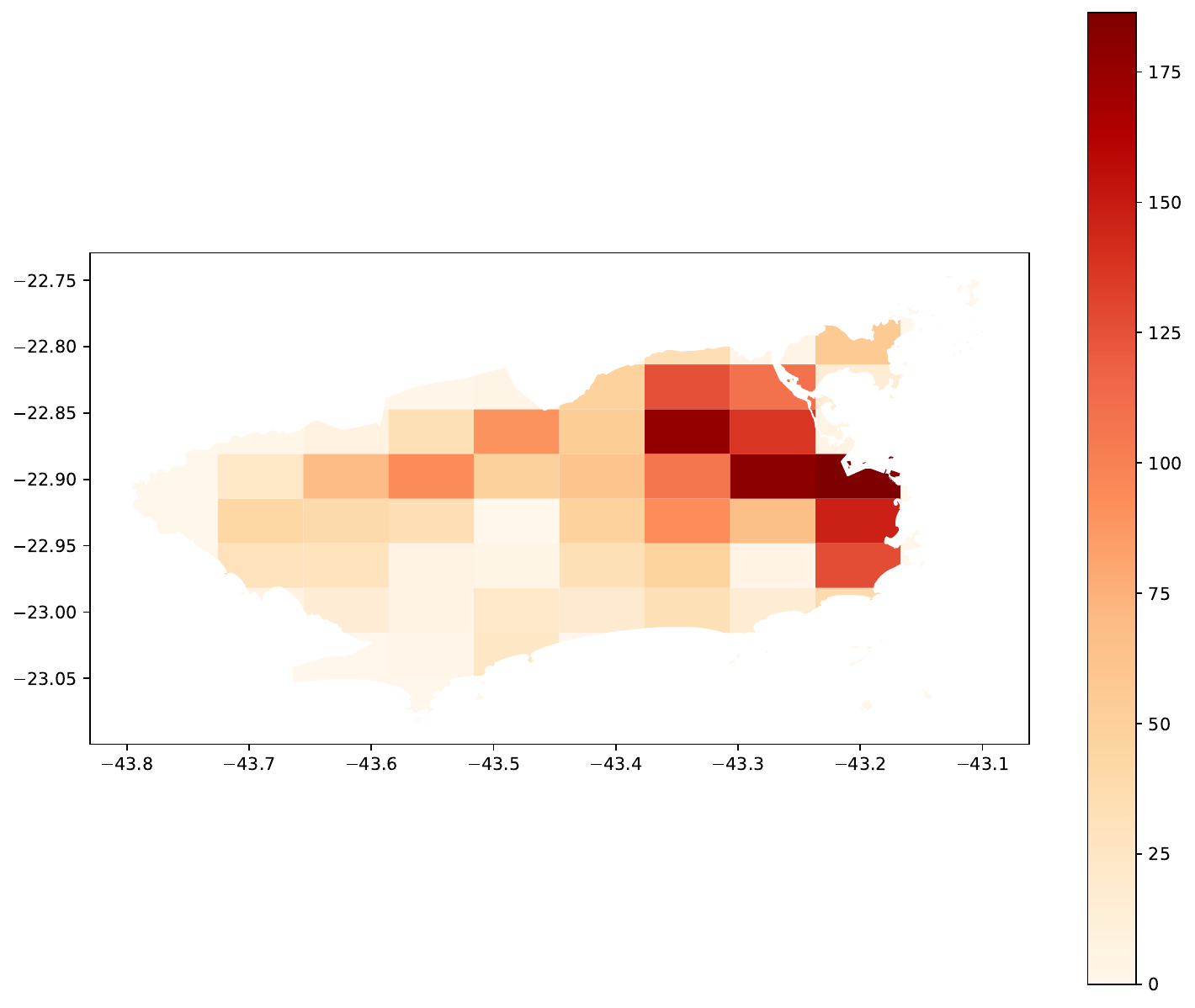} & 
    \includegraphics[scale=0.3]{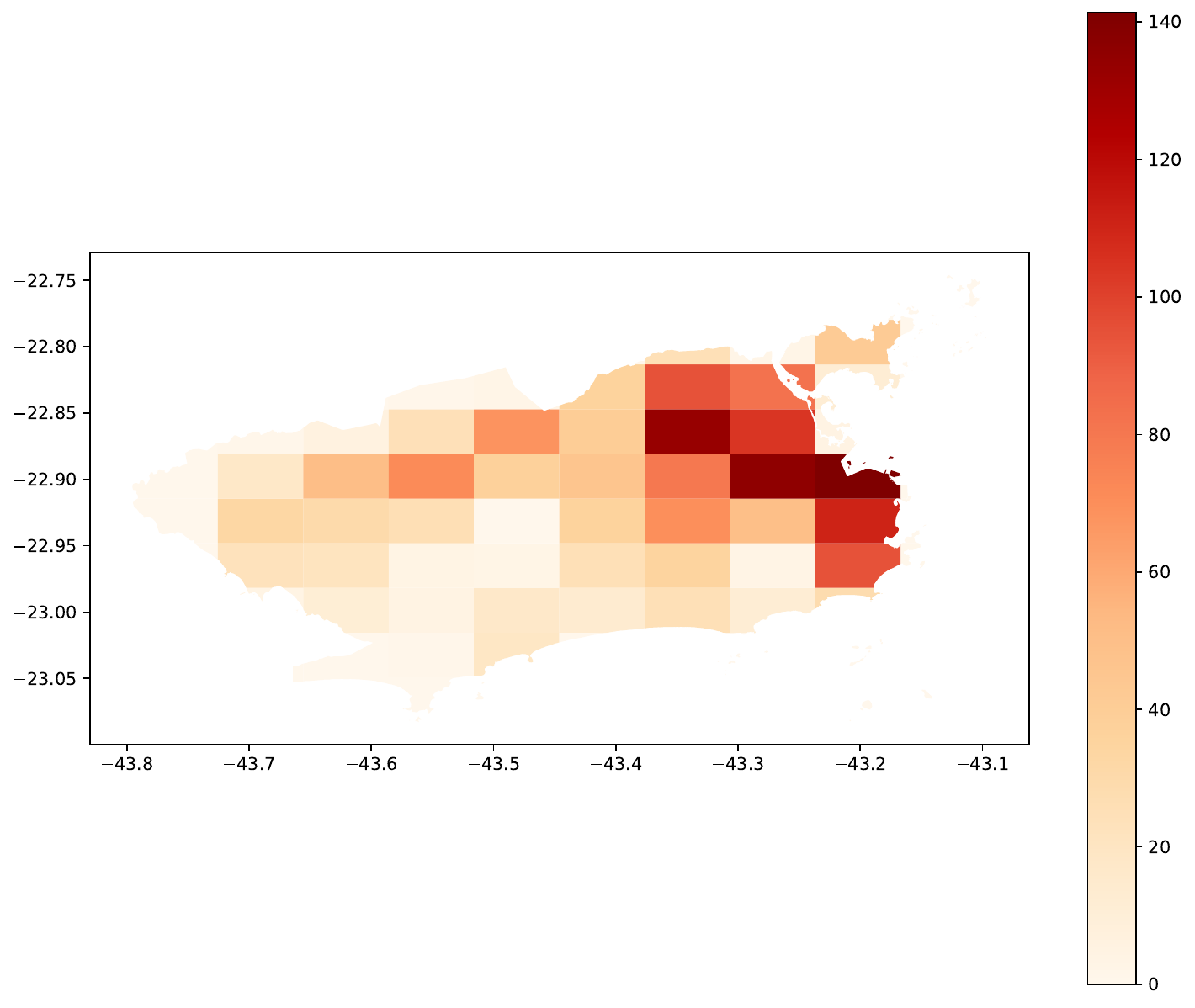}
     \end{tabular}
\caption{Heatmap of the Poisson intensity estimates using data with missing locations (left plots) and without using data with missing locations (right plots) on a rectangular 10x10 discretization.
Top plots: all emergencies.
Bottom plots: high priority emergencies.}
\label{heat1model0}
\end{figure}

\begin{figure}
    \centering
    \begin{tabular}{cc}
        \includegraphics[scale=0.3]{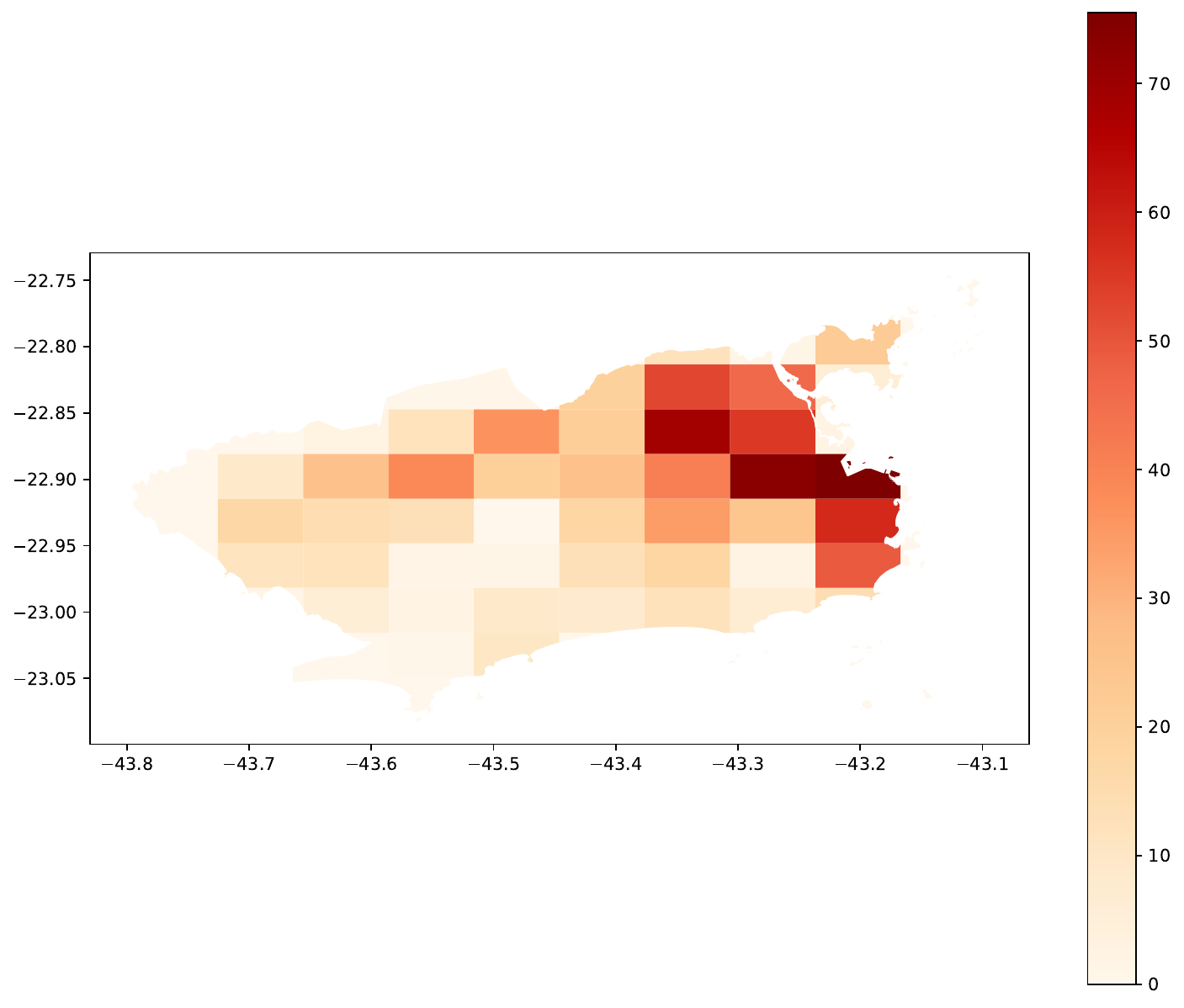} & \includegraphics[scale=0.3]{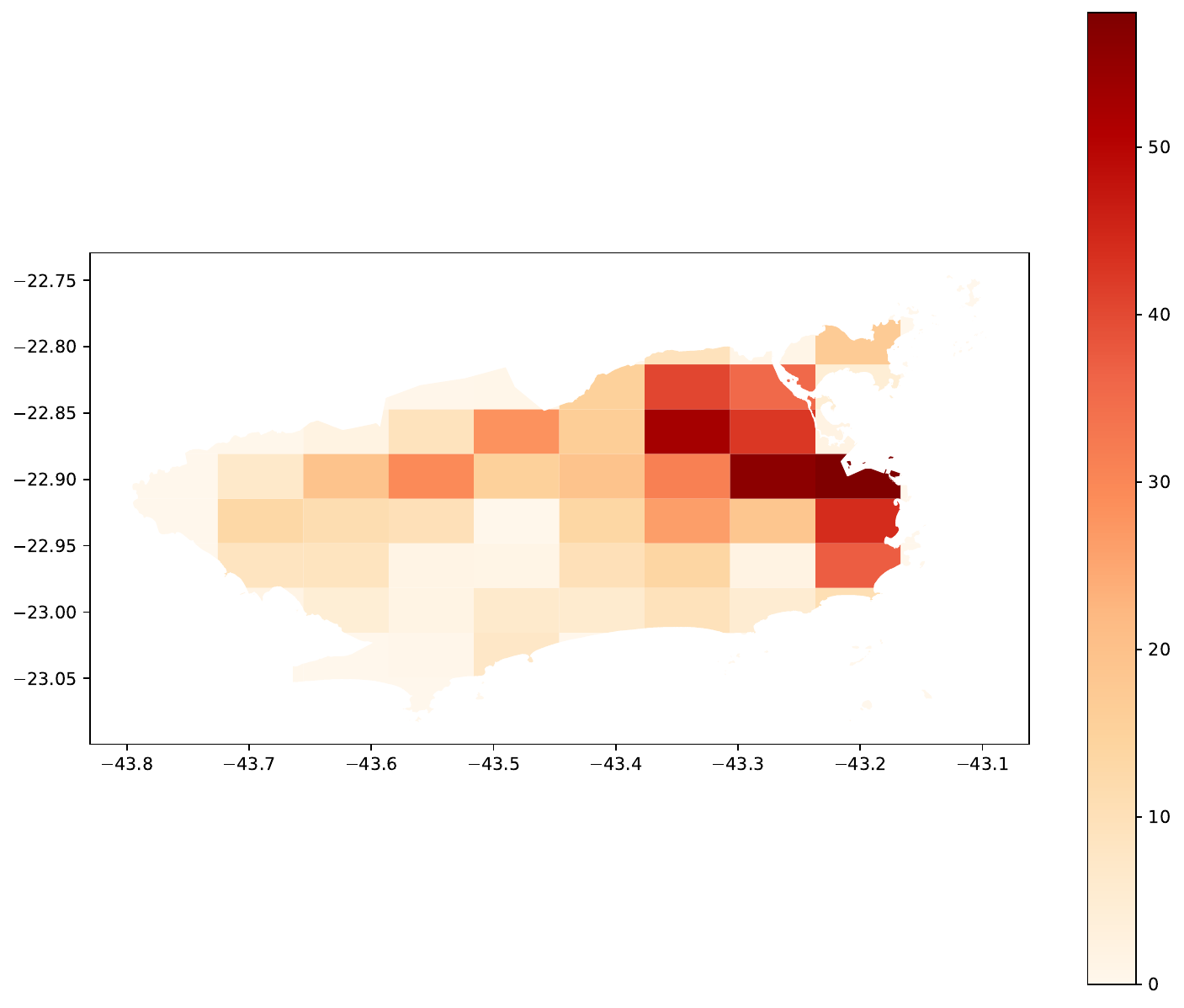}\\     
    \includegraphics[scale=0.3]{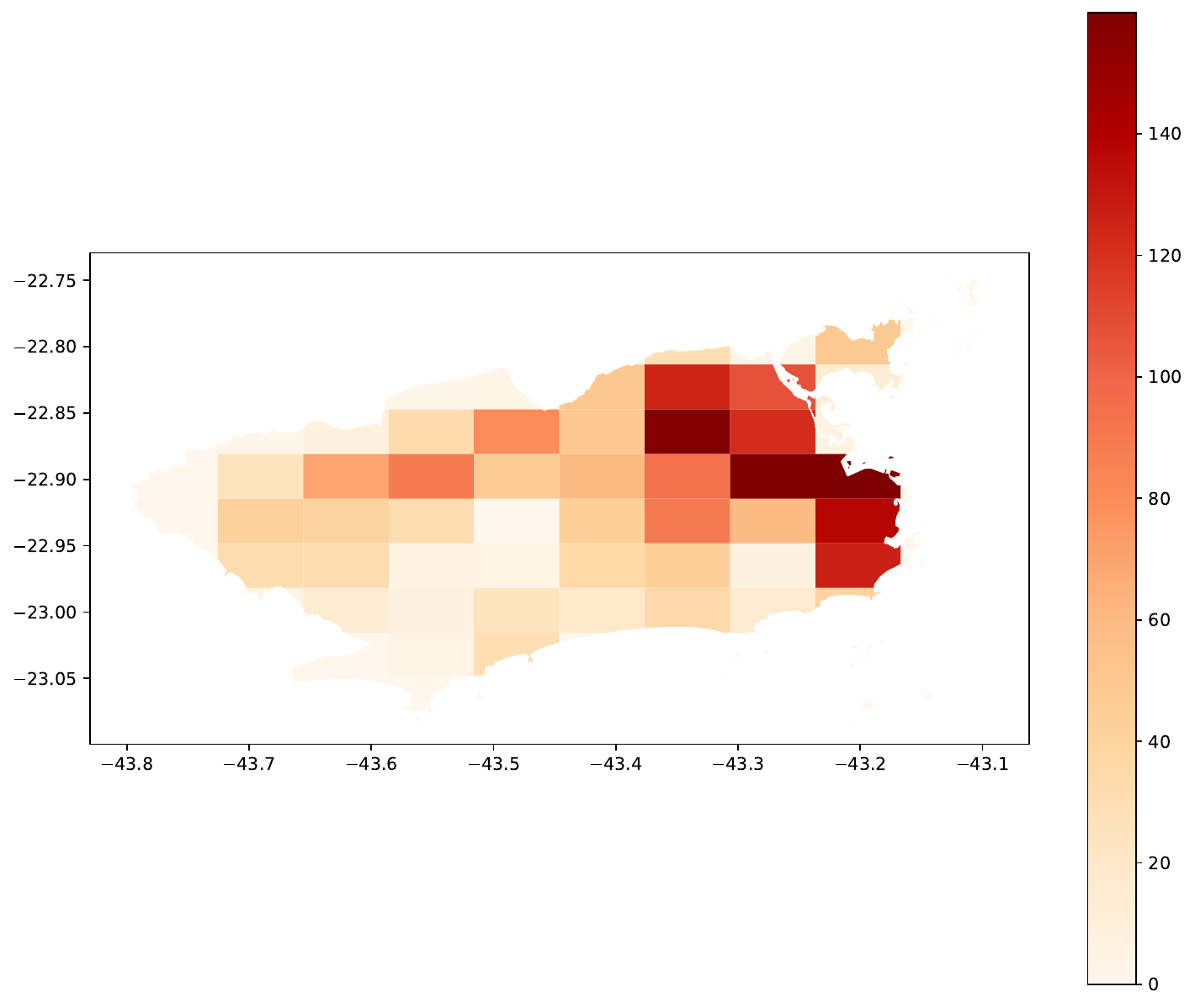} & 
    \includegraphics[scale=0.3]{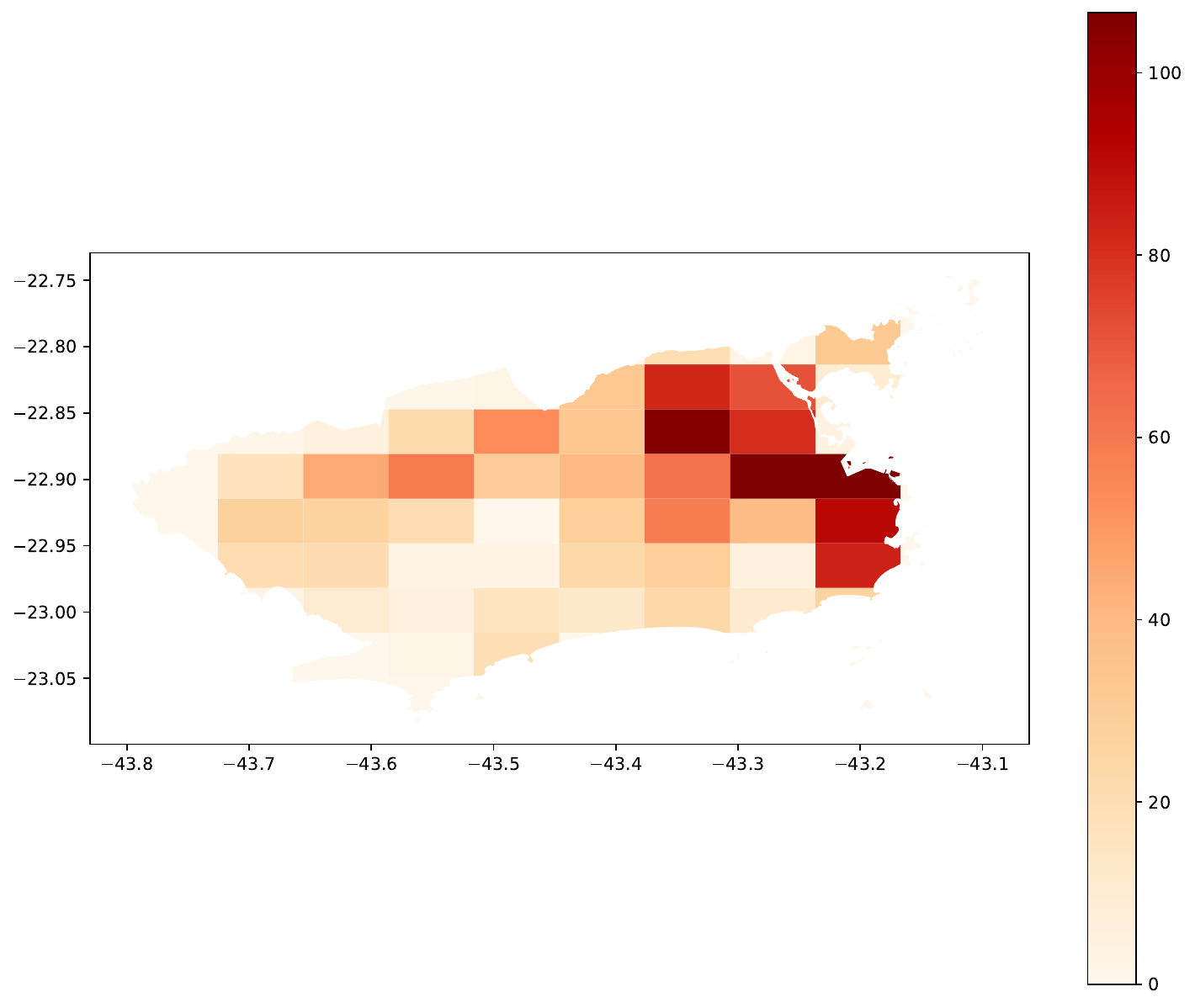}
     \end{tabular}
\caption{Heatmap of the Poisson intensity estimates using data with missing locations (left plots) and without using data with missing locations (right plots) on a rectangular 10x10 discretization.
Top plots: intermediate priority emergencies.
Bottom plots: low priority emergencies.}
\label{heat1model1}
\end{figure}

\section{Replication script}

We provide a replication script that reproduces the numerical results of Section~\ref{sec:num}.
The script and input data are available in the ``Missing\_Data'' subdirectory of the LASPATED replication GitHub repository {\url{https://github.com/vguigues/LASPATED_Replication}}.
The source code and data can be downloaded by either selecting ``Download ZIP'' on the GitHub page, or by installing Git and running:

\begin{verbatim}
    git clone https://github.com/vguigues/LASPATED_Replication.git
\end{verbatim}

The replication script is written in Python and invokes the C++ implementation of the models using the LASPATED library.
The C++ code uses the Boost library ({\url{https://boost.org}}) and the Xtensor and Xtl libraries (both available at {\url{https://github.com/xtensor-stack}}).
We provide a Docker container that handles these dependencies and installs the LASPATED libraries.
To build the Docker container, go to the LASPATED root directory and run:

\begin{verbatim}
    docker build --build-arg USE_GUROBI=0 MISSING_DATA=1 -t laspated .
\end{verbatim}

This command will build a container named laspated.
LASPATED has an optional dependency on the Gurobi optimization library \citep{gurobi}, and the argument USE\_GUROBI=0 disables its use.
The parameter MISSING\_DATA enables the compilation of the C++ code for eatimation with missing data.
After building the container, you can run it in interactive mode:

\begin{verbatim}
    docker run -it laspated
\end{verbatim}

This will open the container as a Linux terminal session.
Next, go to the ``Missing\_Data'' subdirectory and run:

\begin{verbatim}
    cd Missing_Data
    python replication_script.py
\end{verbatim}

This script will invoke the C++ implementation for the models with missing data and generate the output data used in the plots of this paper.
Table~\ref{tab:replication_files} displays the correspondence between the files generated by the script and the figures in Section~\ref{sec:num}.

\begin{table}[]
    \centering
    \begin{tabular}{|c|c|}
    \hline
     results/p\_model1.txt          &  Figure~\ref{empprobmodel1}\\ \hline
     results/lambda\_model1.txt     &  Figures~\ref{emplammodel1},~\ref{histmodel1},~\ref{heat1model0} and~\ref{heat1model1} \\\hline
     results/lambda\_model2wX.txt   &  Figure~\ref{regularized_model} \\\hline
    \end{tabular}
    \caption{Replication files and corresponding figures.}
    \label{tab:replication_files}
\end{table}

The models can be run independently, by executing any of the following commands from the ``Missing\_Data'' subdirectory inside the Docker terminal:

\begin{verbatim}
    ./missing -f test.cfg --model analytical
    ./missing -f test.cfg --model regularized --test_weights 0 0.001 0.005 0.01 0.03
    ./missing -f test.cfg --model population
\end{verbatim}

The first command runs the model of Section~\ref{sec:model1ml}.
The second command runs the model of Section~\ref{sec:reg}, with each of the weights $w$ specified by the ``test\_weights'' argument.
The third command runs the model of Section~\ref{sec:model1m2}.
(The script for the model of Section~\ref{sec:model1m2} can take several hours to run.)
The file test.cfg, specified in the -f option, is a configuration file with some options for the missing data code:

\begin{itemize}
    \item EPS: tolerance for feasibility purposes;
    \item sigma: \(\sigma\) parameter for projected gradient, see \cite{iusem2003, laspatedpaper,laspatedmanual},
    \item max\_iter: maximum number of iterations for projected gradient,
    \item lower\_lambda: lower bound on \(\lambda\) variables,
    \item beta\_bar: initial step size for projected gradient, see \cite{iusem2003, laspatedpaper,laspatedmanual},
    \item test\_weights: list of test weights for the regularized estimator,
    \item model: name of the model to execute, or all to run all models,
    \item info\_file: path to the file describing general information about the model. In the first line, it contains: \(T\), \(D\), \(|\mathcal{I}|\), \(|\mathcal{C}|\) followed by two values that are ignored in this work. \(T\) denotes the number of time intervals during the day of the time discretization (48 for our numerical examples with 48 time windows of 30 mins every day), \(D\) denotes the number of days in a period ($7$ in our numerical examples). The second line contains the number of observations for each day of the week, starting on Monday;
    \item arrivals\_file: path to the file describing the samples for which the location \(i \in \mathcal{I}\) is known. Each line denotes a sample \(M_{c,i,t,n}^{1}\) and contains the time interval index values~$t$ (in our numerical examples, $t \in \{1,\ldots,48\}$, with $30$~minute time intervals), day index values~$d$ (in our numerical examples, $d \in \{1,\ldots,7\}$), zone index~$i$, arrival type index $c$, observation index~$n$, \(M_{c,i,t,n}^{1}\), and a value $h$ that is ignored in this work (indicates in LASPATED whether the day is holiday or not).
    \item neighbors\_file: path to the file describing the zones. Each line contains the zone index~$i$, the latitude and longitude of its center, the zone type (not used for the models of this paper, see \cite{laspatedmanual,laspatedpaper} for the use of the zone type in LASPATED), five features about the zone and a list of neighbors of~$i$ and its corresponding distances. 
    \item missing\_file: path to the file containing samples where the zone is unknown. The format is similar to arrrivals\_file, but the zone index is ignored when reading this file.
\end{itemize}

The options may also be passed via command line arguments, and the command line options take precedence over the same options written in the configuration file.

\section{Conclusion}
\label{sec:conc}

We presented four types of models to deal with missing locations in spatiotemporal data.
The models were implemented as an extension of the LASPATED software available at {\url{https://github.com/vguigues/LASPATED/Missing_Data}}.
Numerical results were presented using the history of emergency calls for the Rio de Janeiro emergency medical service, where the locations of many emergencies are not recorded.

\addcontentsline{toc}{section}{References}
\bibliographystyle{plain}
\bibliography{Biblio}
\if{

\begin{figure}
\centering
\begin{tabular}{c}
\includegraphics[scale=0.8]{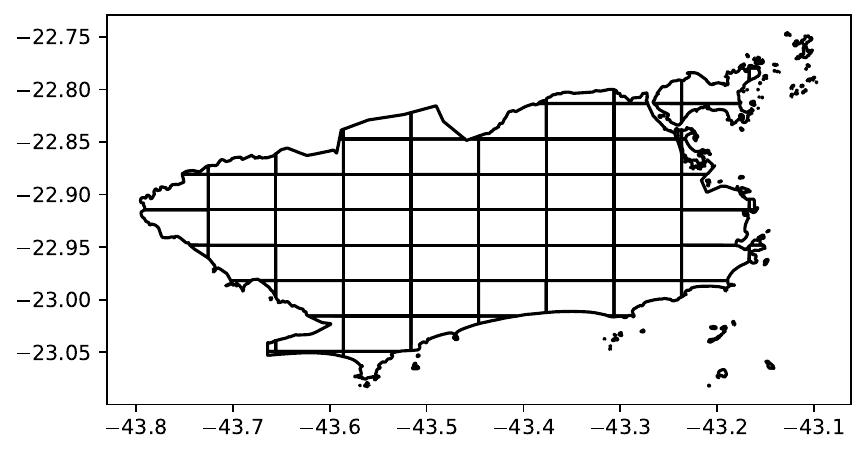}
\end{tabular}
\caption{\label{figurerect1010}
Space discretization of a region containing the city of Rio de Janeiro into $10 \times 10 = 100$ rectangles, 76 of which have nonempty intersection with the region and are shown in the figure.}
\end{figure}

\begin{figure}
    \centering
    \begin{tabular}{c}
        \includegraphics[scale=0.9]{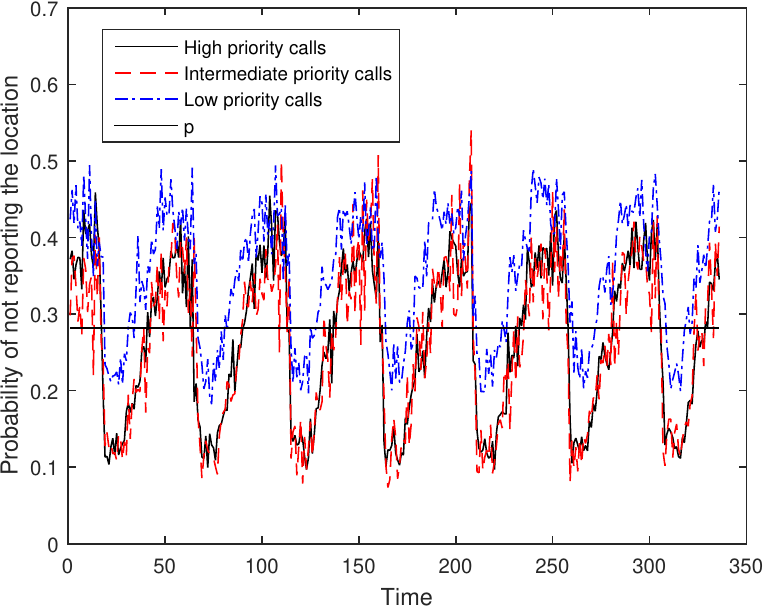}        
     \end{tabular}
 \caption{Empirical probabilities of not reporting the location for calls of high priority, of intermediate priority, and of low priority.
 We also report the estimation of $p$ for a model where the probability of not reporting the call does not depend on time and call priority.}
 \label{empprobmodel1}
\end{figure}

\begin{figure}[hbtp]
    \centering
    \begin{tabular}{cc}
        \includegraphics[scale=0.60]{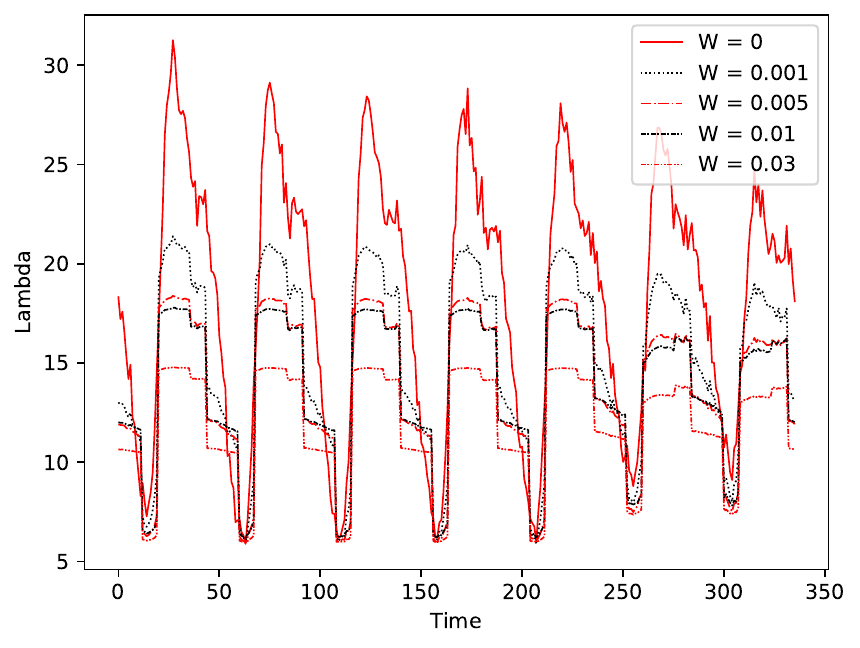} & \includegraphics[scale=0.60]{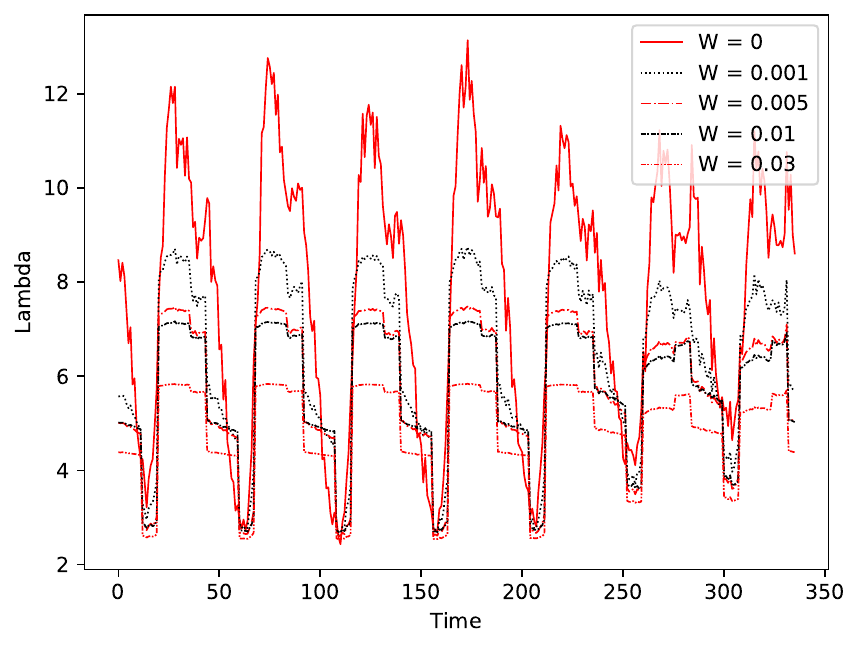}\\     
        \includegraphics[scale=0.60]{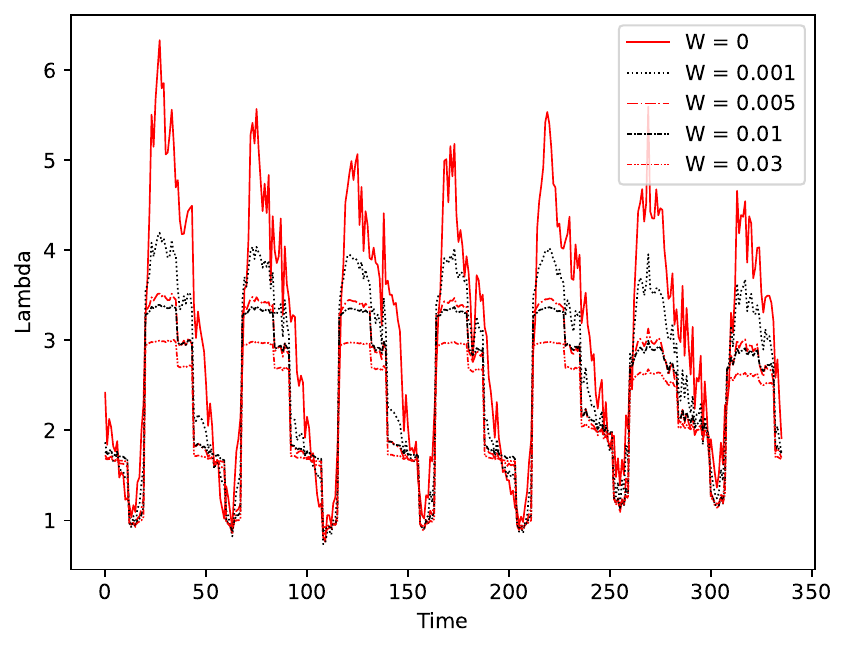} & \includegraphics[scale=0.60]{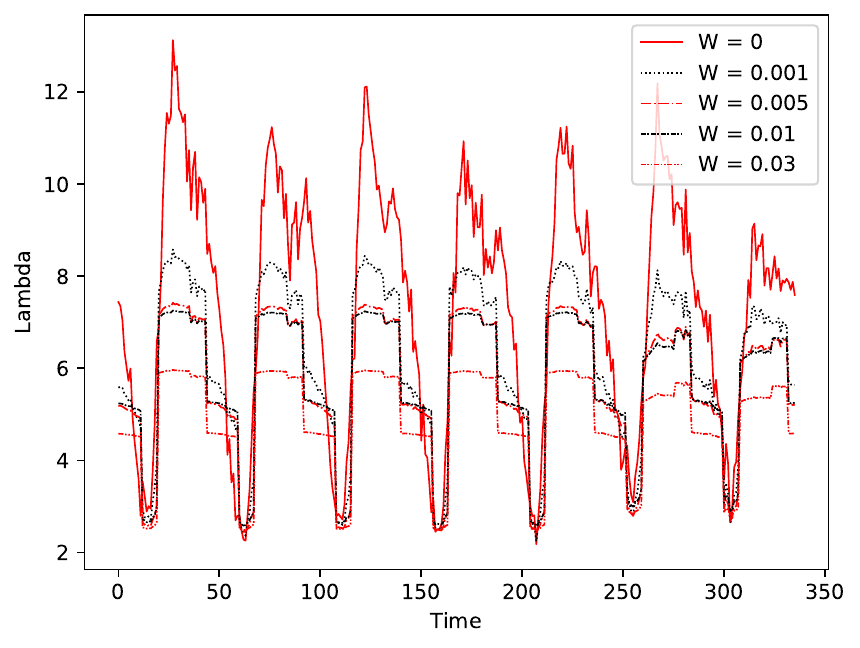}
     \end{tabular}
 \caption{Poisson intensities for the regularized model with different weight values $w$. The top-left figure is the sum of intensities over all priorities, the top-right contains the intensities for high-priority calls, the bottom-left displays the intensities for intermediate-priority calls, and the bottom-right plot corresponds to low-priority calls intensities.}
 \label{regularized_model}
\end{figure}

\begin{figure}
    \centering
    \begin{tabular}{cc}
        \includegraphics[scale=0.63]{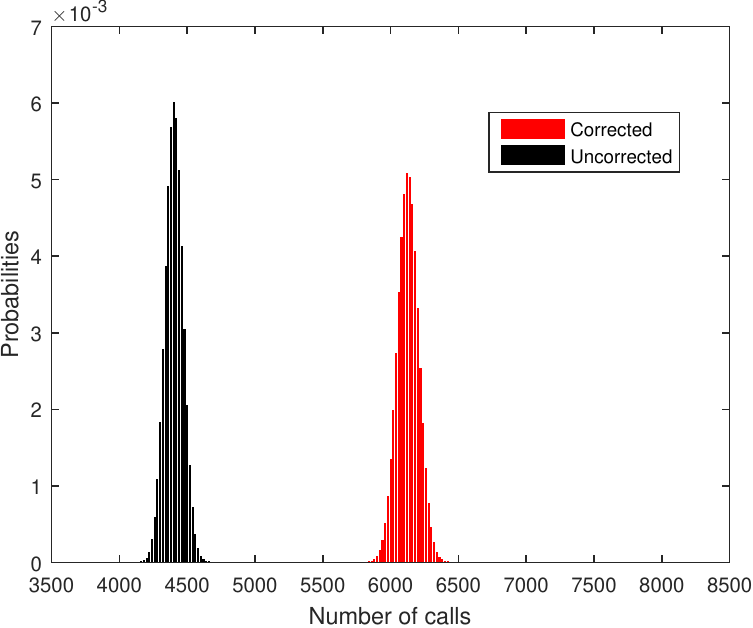} & \includegraphics[scale=0.63]{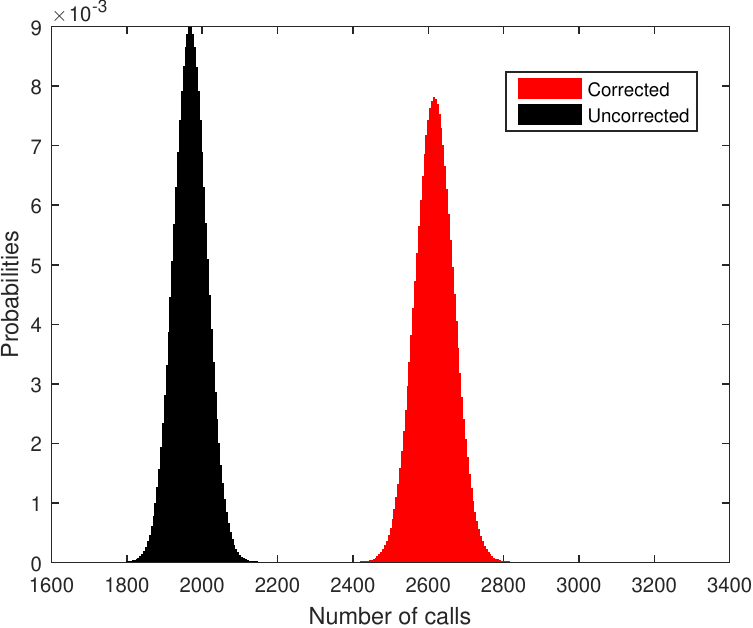}\\     
        \includegraphics[scale=0.63]{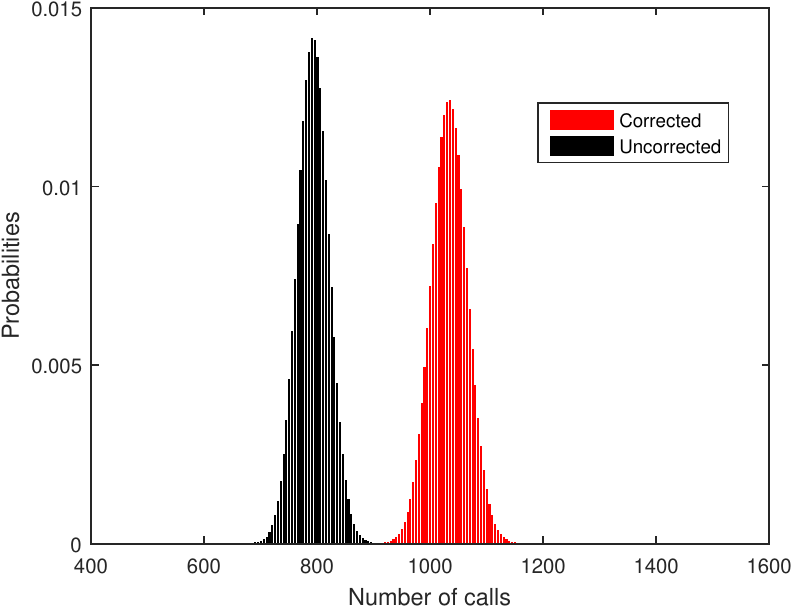} & \includegraphics[scale=0.63]{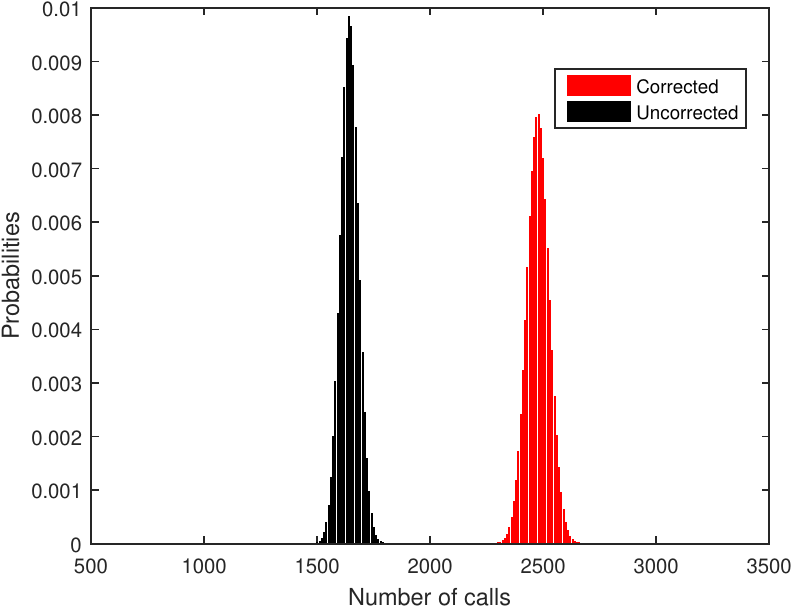}
     \end{tabular}
 \caption{Histogram of the distribution of the number of calls
 over all regions and for the whole week
 considering calls with missing locations (in red) and without considering these calls (in black).
 Top left: all priorities, top right: calls of 
 high priority, bottom left: calls of intermediate
 priority, bottom right: calls of low priority.}
 \label{histmodel1}
\end{figure}

\begin{figure}
    \centering
    \begin{tabular}{cc}
        \includegraphics[scale=0.65]{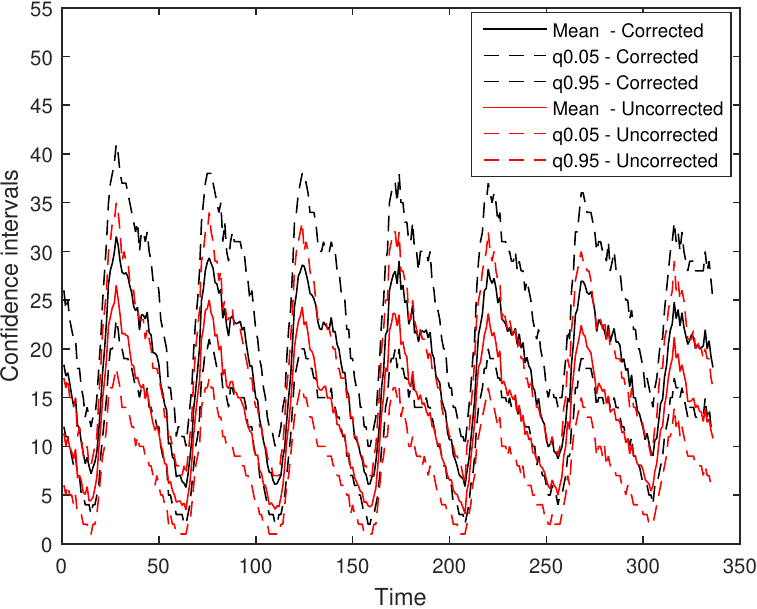} & \includegraphics[scale=0.65]{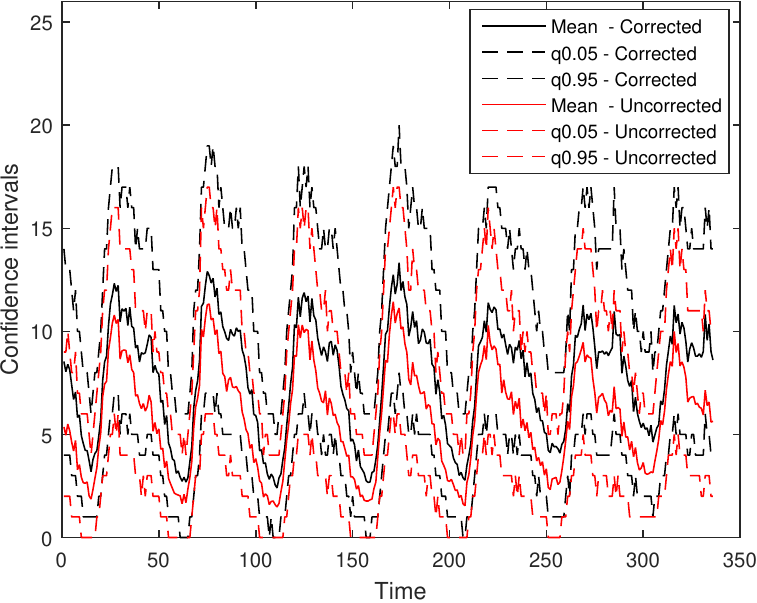}\\     
        \includegraphics[scale=0.65]{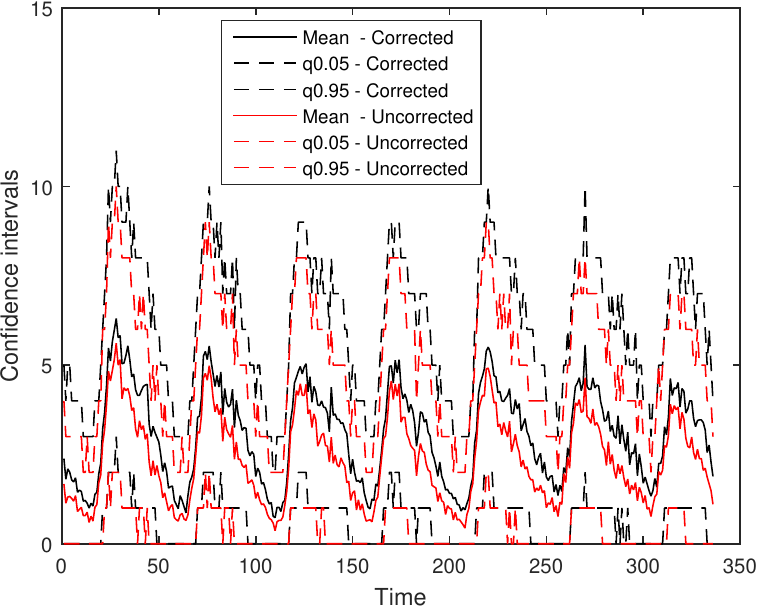} & \includegraphics[scale=0.65]{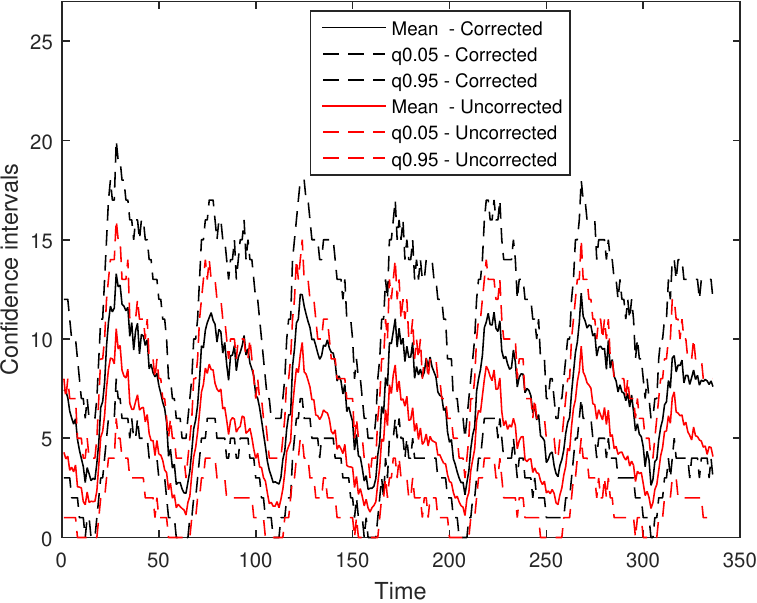}
     \end{tabular}
 \caption{Confidence intervals on the number of calls along the time intervals of 30 mins during the week.
 Top left: all calls. Top right: calls of high priority.
 Bottom left: calls of intermediate priority.
 Bottom right: calls of low priority.}
 \label{emplammodel1}
\end{figure}

\begin{figure}
    \centering
    \begin{tabular}{cc}
        \includegraphics[scale=0.3]{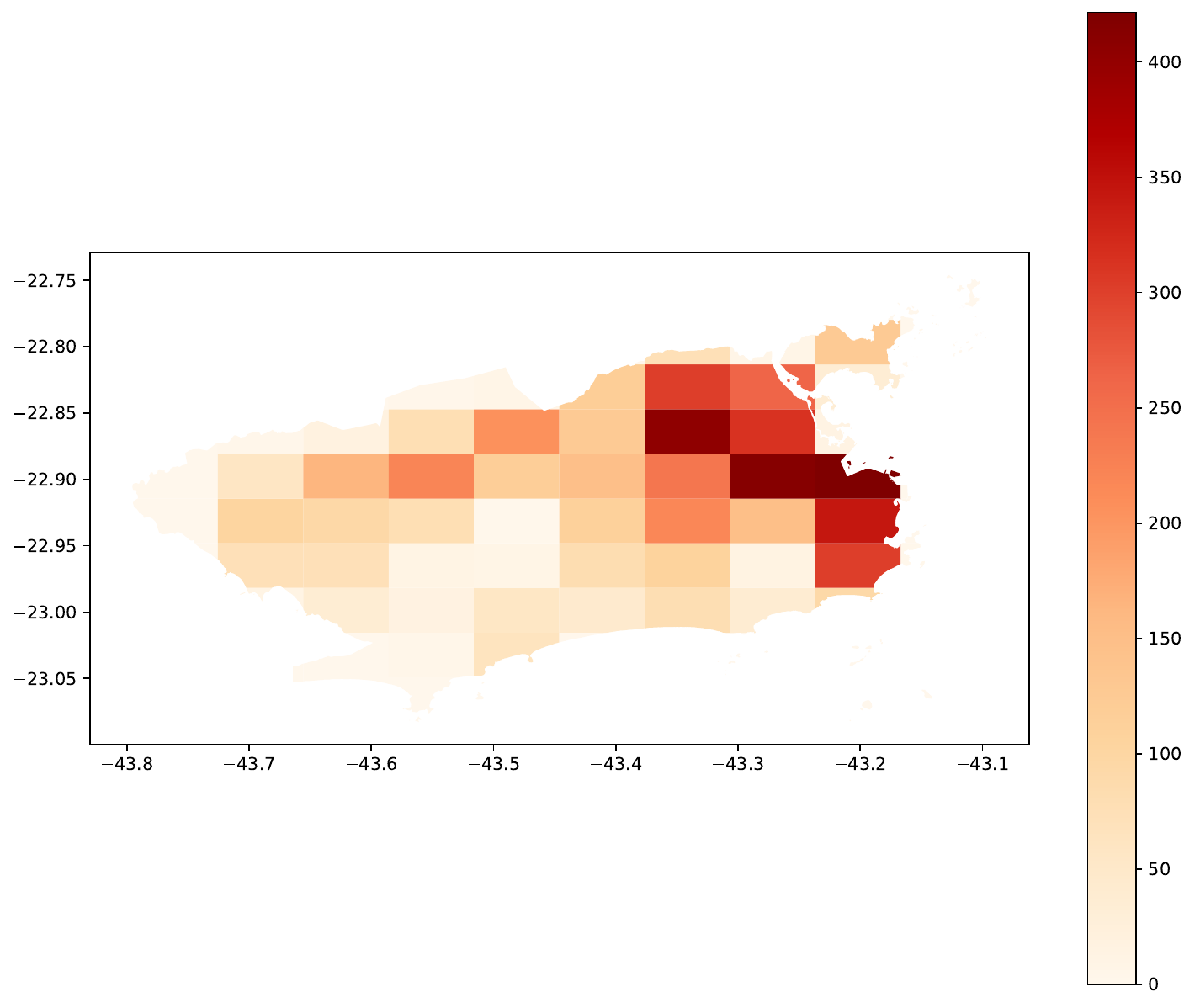} & \includegraphics[scale=0.3]{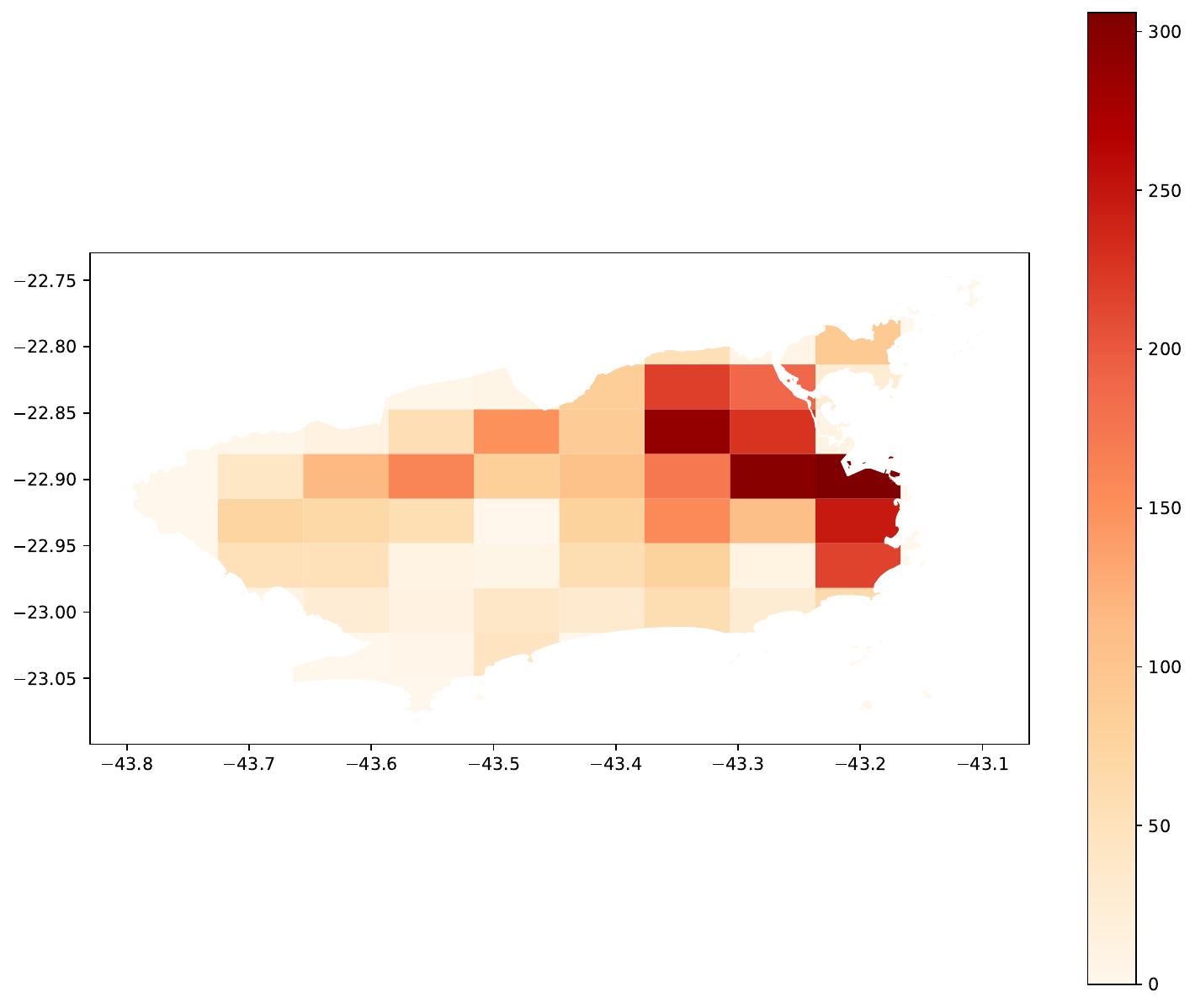}\\     
    \includegraphics[scale=0.3]{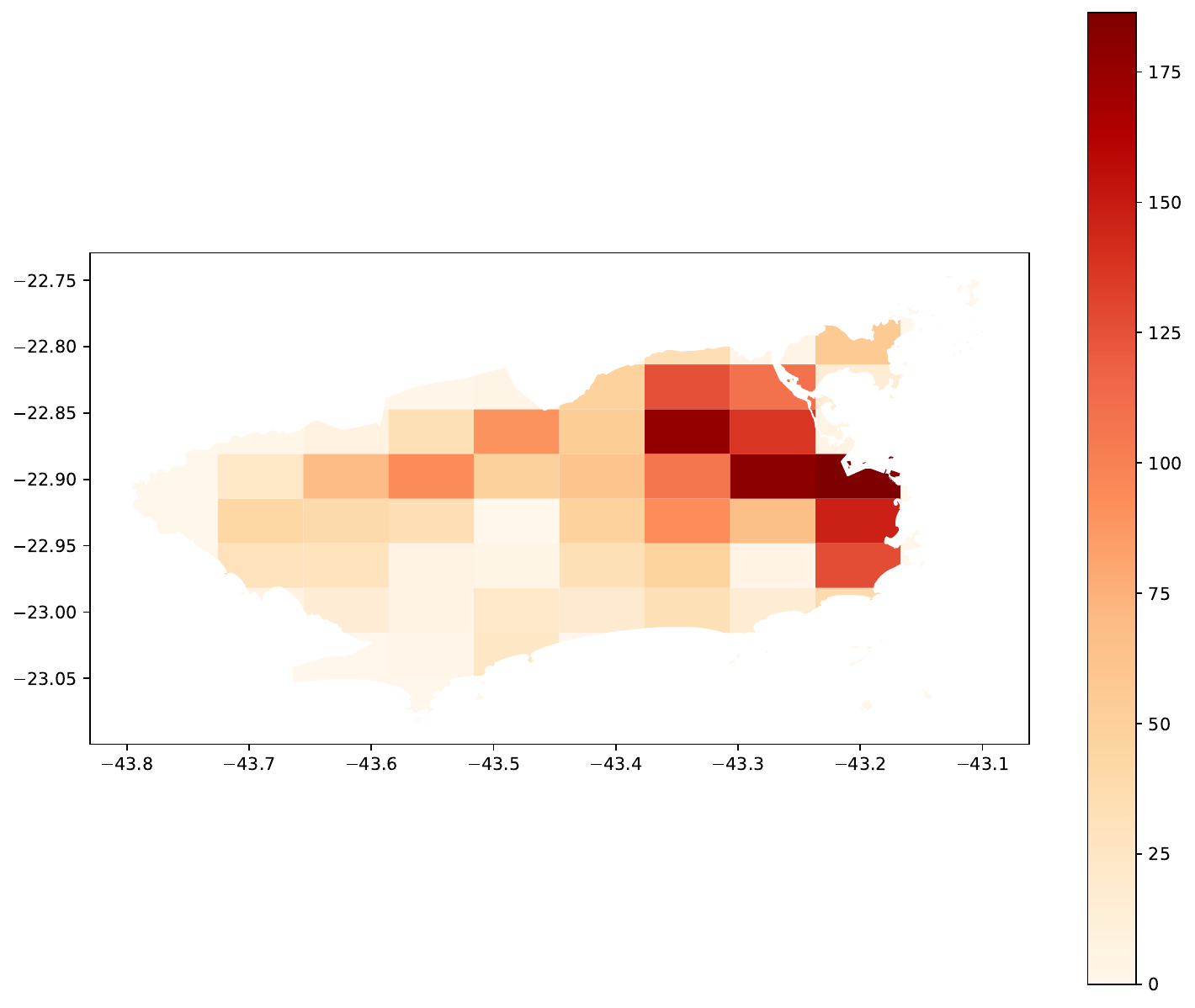} & 
    \includegraphics[scale=0.3]{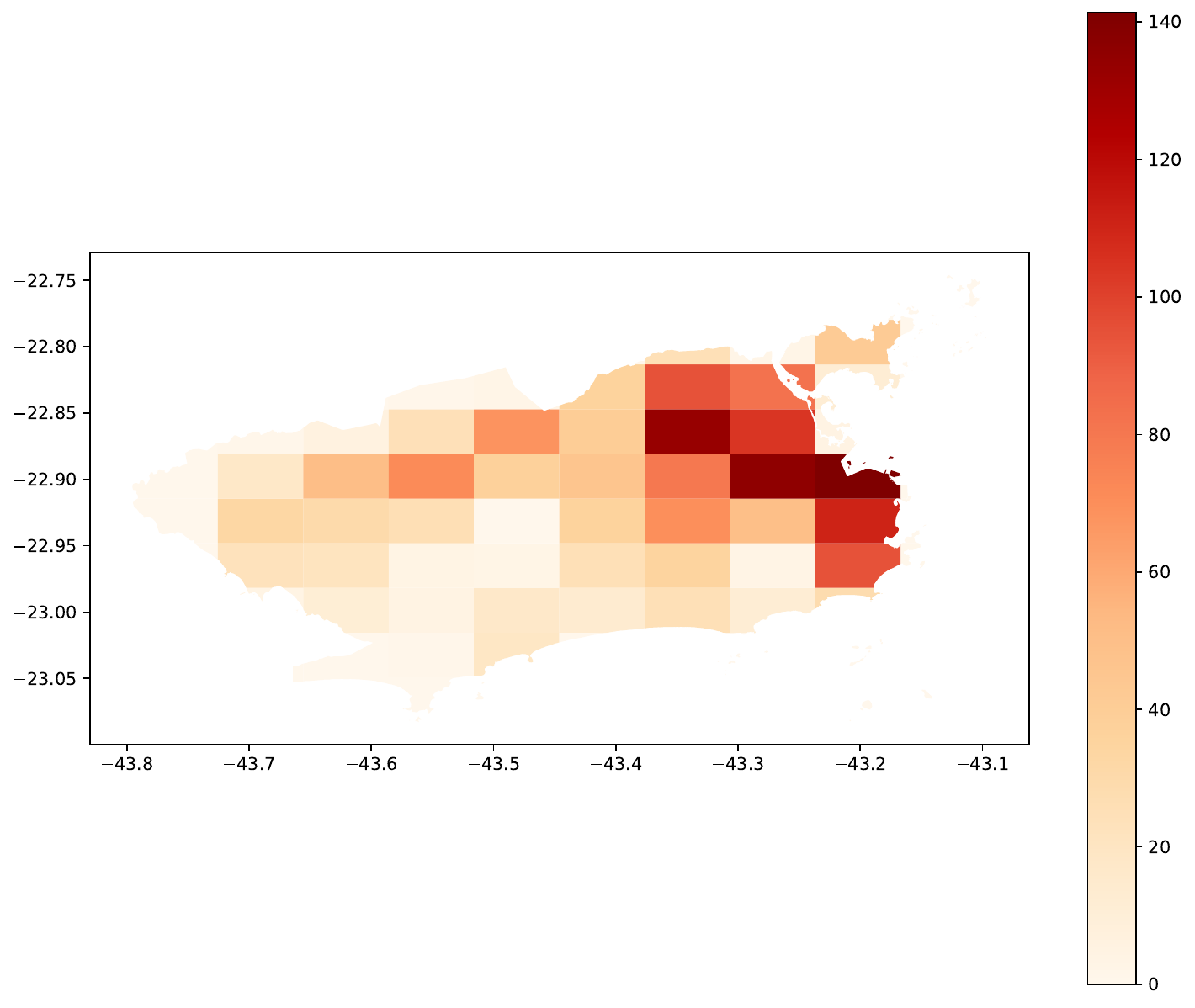}
     \end{tabular}
 \caption{Heatmap of the Poisson intensities considering missing locations (left plots)
 and discarding calls with missing locations (right plots) and a rectangular 10x10 discretization. 
 Top plots: all calls. Bottom plots: Calls of high priority.}
 \label{heat1model0}
\end{figure}

\begin{figure}
    \centering
    \begin{tabular}{cc}
        \includegraphics[scale=0.3]{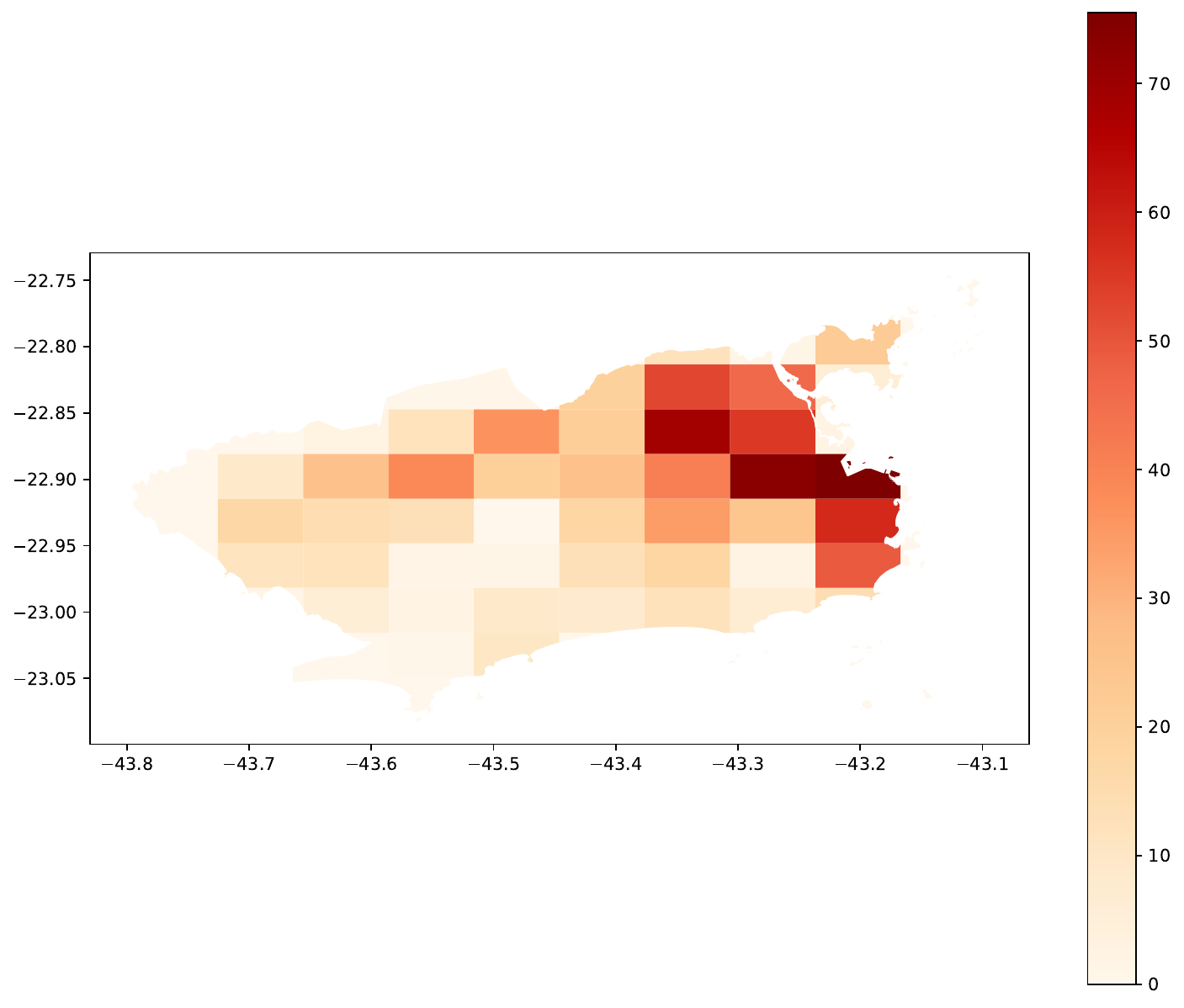} & \includegraphics[scale=0.3]{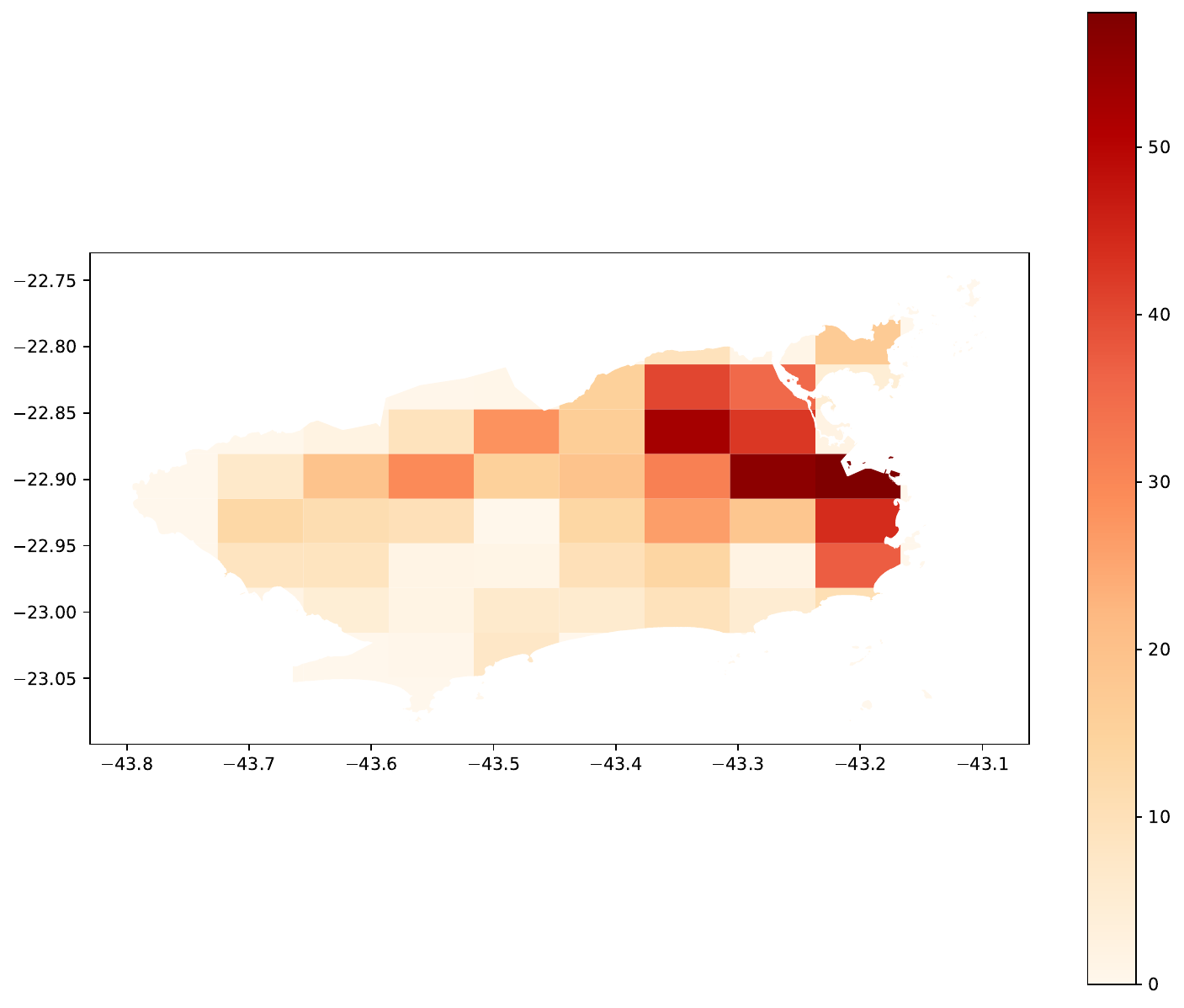}\\     
    \includegraphics[scale=0.3]{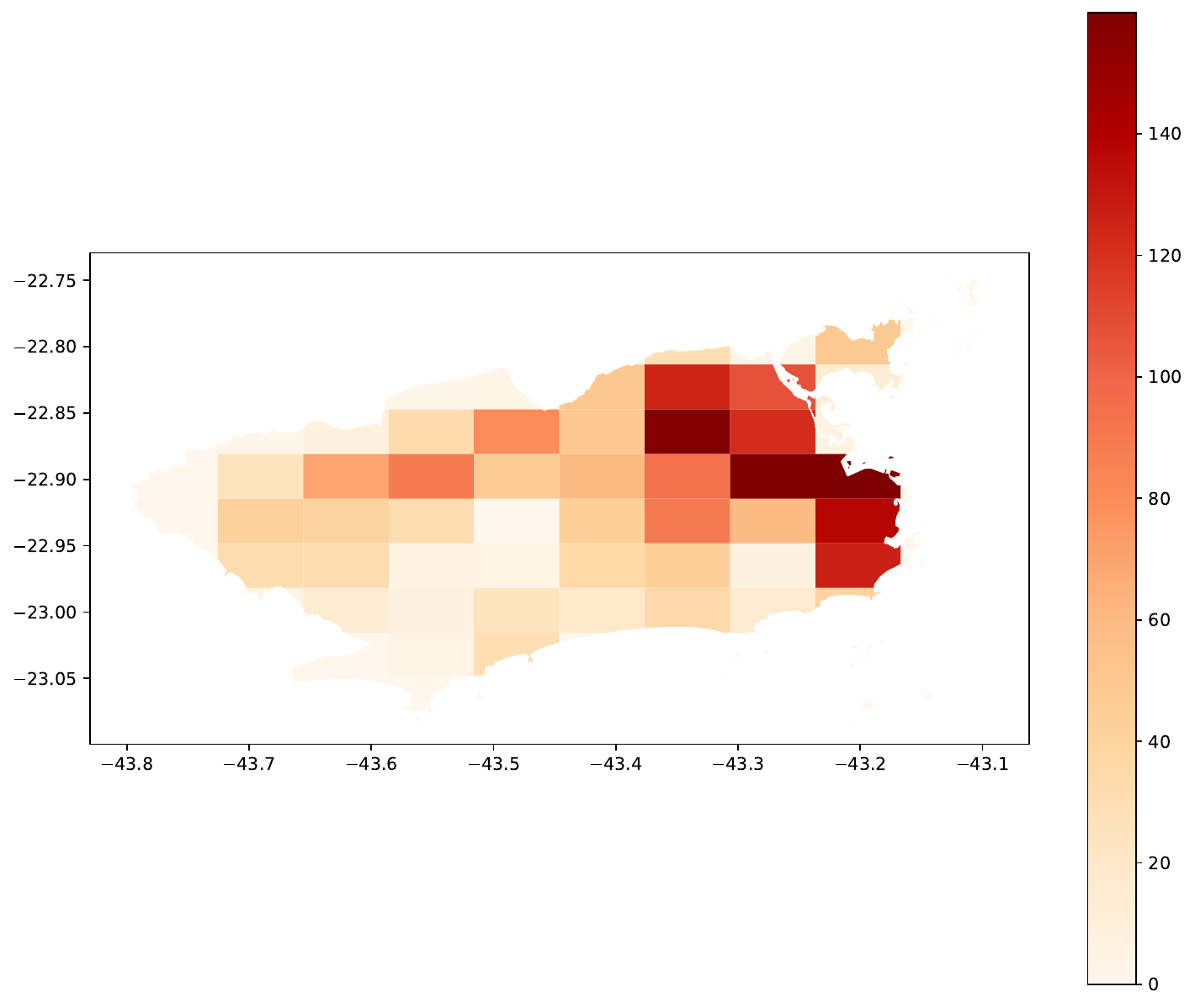} & 
    \includegraphics[scale=0.3]{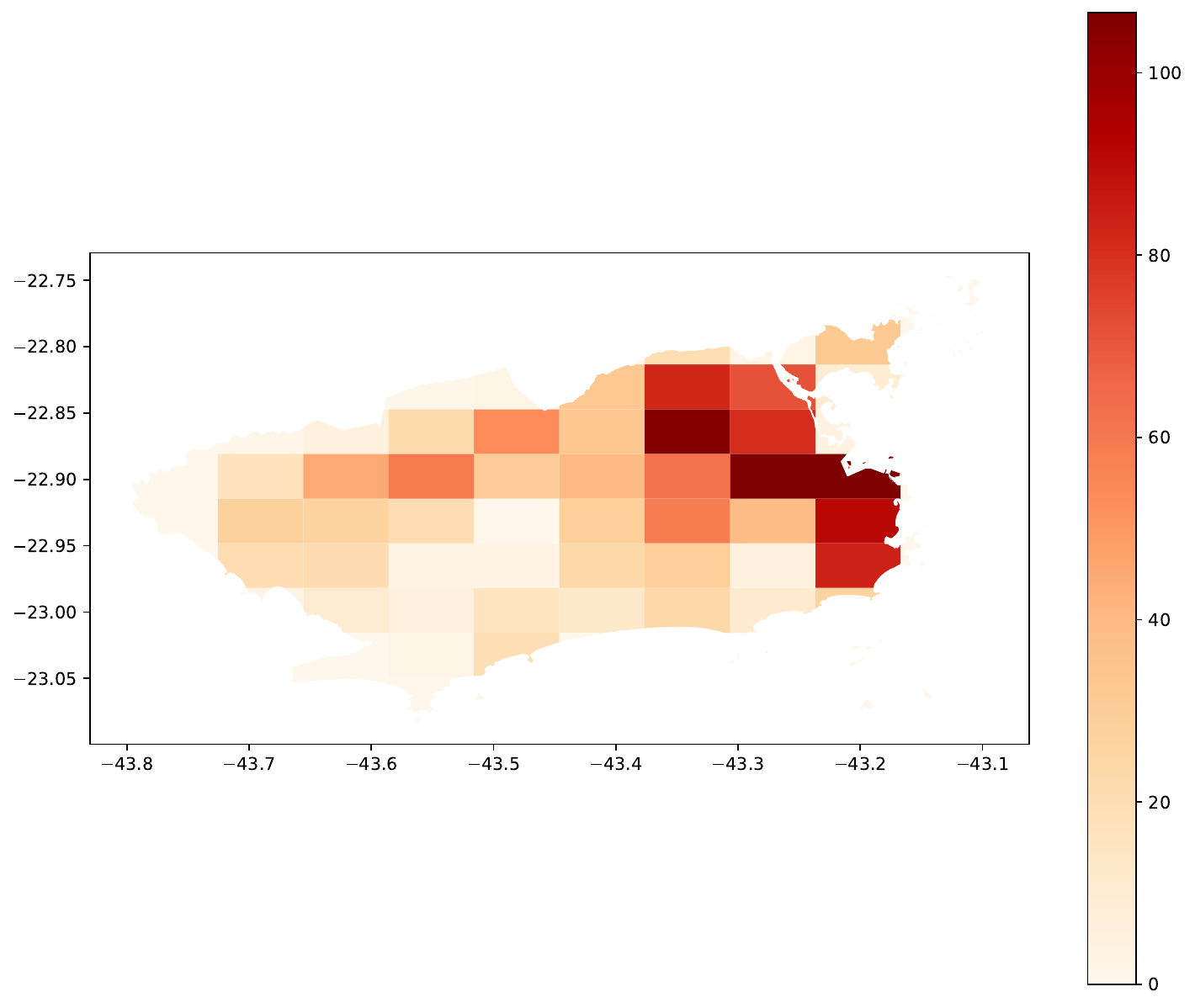}
     \end{tabular}
 \caption{Heatmap of the Poisson intensities considering missing locations (left plots)
 and discarding calls with missing locations (right plots) and a rectangular 10x10 discretization. 
 Top plots: calls of intermediate priority. Bottom plots: Calls of low priority.}
 \label{heat1model1}
\end{figure}

\begin{figure}
\centering
\begin{tabular}{c}
\includegraphics[scale=0.8]{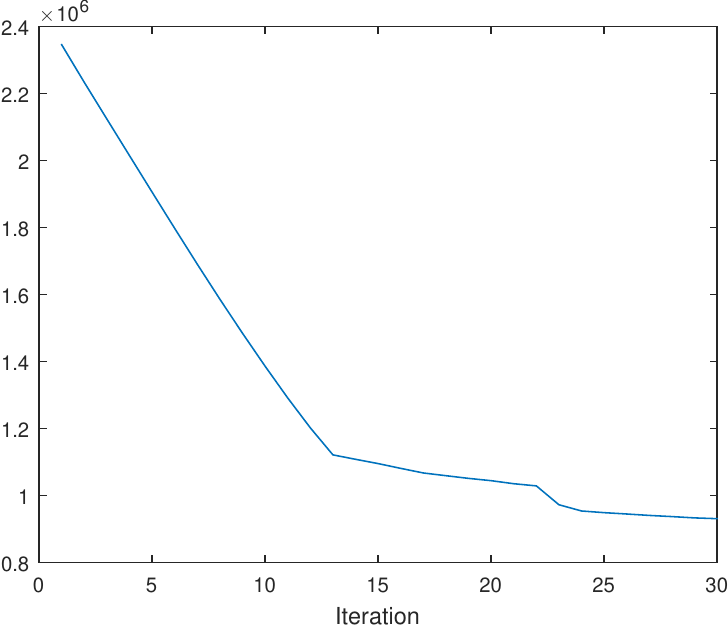}
\end{tabular}
\caption{\label{figurepg}
Evolution of the approximate
optimal values along iterations of projected gradient
with line search along the feasible direction, applied to problem
\eqref{pboptinitr} with all penalizations
$W_{G}$ and $w_{i,j}$
equal to one.}
\end{figure}

}\fi

\end{document}